\documentclass[12pt,preprint,xdvi]{aastex}

\usepackage{psfig}

\newcommand{\be}{\begin{equation}}
\newcommand{\ee}{\end{equation}}
\newcommand\eq{Equation}

\newcommand\fig{Figure}
\newcommand\figs{Figures}
\newcommand\etal{{\it et al.}}

\newcommand{\avg}[1]{{\langle{#1}\rangle}}

\newcommand{\xvec}{{\bf x}}
\newcommand{\kvec}{{\bf k}}

\newcommand{\xhat}{{\bf \hat{x}}}
\newcommand{\yhat}{{\bf \hat{y}}}
\newcommand{\papi}{LBP03}

\begin{document}
%\begin{article}
%\begin{opening}

\title{The Number of Magnetic Null Points in the Quiet Sun Corona}

\author{D.W. Longcope,$^1$ C.E. Parnell$^2$}
\affil{1. Department of Physics, Montana State University\\
  Bozeman, Montana 59717\\
2. School of Mathematics and Statistics, 
University of St Andrews,\\St Andrews, Fife, UK, KY16 9SS}


\begin{abstract}
The coronal magnetic field above a particular photospheric region
will vanish at a certain number
of points, called null points.  These points can be found
directly in a potential field extrapolation or their density 
can be estimated from Fourier spectrum of the magnetogram.
The spectral estimate, which
assumes that the extrapolated field is random, homogeneous and
has Gaussian statistics, is found here to be relatively accurate for 
quiet Sun magnetograms from SOHO's MDI.  The majority of null points
occur at low altitudes, and their
distribution is dictated by high wavenumbers in the Fourier spectrum.
This portion of the spectrum is affected by Poisson
noise, and as many as five-sixths of null points
identified from a direct extrapolation can be attributed to noise.
The null distribution above 1500 km is found to depend on wavelengths that
are reliably measured by MDI in either its low-resolution or 
high-resolution mode.  After correcting the spectrum to remove
white noise and compensate for the modulation transfer function
we find that a potential field extrapolation contains, on
average, one magnetic null point, with altitude greater than 
1.5 Mm, above every $322\,{\rm Mm}^2$ patch of quiet Sun.  Analysis of 562
quiet Sun magnetograms spanning the two latest solar minima shows that
the null point density is relatively constant with roughly 10\% 
day-to-day variation.
At heights above 1.5 Mm, the null point density decreases
approximately as the inverse cube of height.
The photospheric field in the quiet Sun is well approximated as that
from discrete elements with mean flux $\avg{|\phi|}=1.0\times10^{19}$ Mx
distributed randomly with density $n=0.007\,{\rm Mm}^{-2}$.
\end{abstract}

\date{Draft: \today}

%\end{opening}

\section{Introduction}

Models of magnetic reconnection, from the very earliest, have assigned
particular significance to points where the magnetic field vanishes:
magnetic null points \citep{Dungey1958,Sweet1958}.  
In two-dimensional reconnection 
models, null points (also called neutral points) are
natural locations to change magnetic field line topology.  While
three dimensions offer alternatives for topological change
\citep{Greene1988,Hesse1988}, null
points are still natural locations for current intensification
\citep{Craig1995,Rickard1996,Galsgaard1997,Pontin2007} 
and for focusing of magnetosonic waves 
\citep{Hassam1992,Craig1993,McLaughlin2004}.  Dissipation of magnetic
energy at magnetic null points is therefore a potential source of
heat in the quiet Sun corona.  The effectiveness of this dissipation
for heating the entire  corona will depend, in part, in how null
points are distributed throughout its field.

% effectiveness of heating by reconnection at coronal null points
%  would seemingly depend on how commonly such structures are found in
%  the coronal field.

% Jets: common features in X-ray observations.  more common w/ Hinode
%  [Shibata models.  Hinode observations]
%   models involve null point at cusp.
%   Are null points common enough to explain the ubiquity of jets?

Coronal jets are taken as another observational manifestation of
coronal null points.  Observations in EUV and X-rays show that jets, 
with their characteristic apex cusp \citep{Shibata1992,Shimojo1996},
appear to be more common  in coronal holes and the quiet Sun 
than previously believed \citep{Culhane2007,Cirtain2007}.  
Theoretical models of jets invoke magnetic reconnection occurring 
at a null point many megameters above the photospheric surface
\citep{Yokoyama1996,MorenoInsertis2007}.  In order that such a model apply
to the large number of jets observed it would be
necessary that magnetic null points are relatively common at the
altitudes of observed cusps.

% Complicated magnetic fields
%   field generally believed to get simpler w/ height:
%   beta decreases, influence of small photospheric structuring
%   diminishes

In many models, coronal null points occur above a photospheric field
consisting of one magnetic element completely surrounded
by a ring of opposing polarity \citep{Antiochos1998,MorenoInsertis2007}.
The null point will persist even if the continuous surrounding ring
is broken, but it remains
necessary that opposing polarity be found on ``all sides'' of the
central element \citep{Seehafer1986,Beveridge2004}.  This suggests
that coronal null points occur only under special circumstances which
are likely to be found infrequently in the actual solar photosphere, and
raises once more the question of how common null points might be
in the actual coronal magnetic field. 

% QS models: Schriver & Title, point sources.
%   LBP03: general theory assuming homogeneous Gaussian field
%   arising from p-spheric spectrum.  Density fell rapidly w/ height.
%   for flat spectrum, such as from white noise or point sources
%     \sim z^{-3}.

One recent investigation by R\'egnier, Parnell and Haynes (2008)
\nocite{Regnier2008} found 80 magnetic null
points over a $102\,{\rm Mm}\times116\,{\rm Mm}$ patch of quiet Sun.  A
potential field was extrapolated onto a rectilinear grid from a
magnetogram made by Hinode's NFI instrument.  The algorithm of
Haynes and Parnell (2007)
\nocite{Haynes2007} was then applied to the gridded field to locate all
points where an interpolated field would vanish.  This revealed that,
at least over that region at that time, there was a column of
$N_n=6.7\times 10^{-3}$ null points per square megameter, about half
above a height of $z=1$ Mm.
Close, Parnell and Priest (2004)
\nocite{Close2004b} extrapolating from high-resolution MDI 
magnetograms of a different day found $1.7\times10^{-3}\,{\rm
Mm}^{-2}$, almost four times fewer.  This discrepancy suggests a
potential dependence on instrument and specific date of observation.
A more general measurement of null point density, applicable to any
time or any region, would require the repetition of this process.

Several investigations have estimated the density of magnetic null
points in generic quiet Sun coronal fields.  Schrijver and Title (2002)
\nocite{Schrijver2002}
modeled the photospheric field of the quiet Sun as a balanced, random
distribution of magnetic point charges.  They found that the potential
field anchored to these charges vanished at one point
in the corona, on average, for every eleven point charges on the
photosphere.  They estimated that the density of null points
diminished exponentially with height.

Longcope, Brown and Priest (2003, hereafter \papi)
\nocite{Longcope2003b} developed a more general approach to estimating
the density of coronal null points in a potential 
coronal field.  They assumed the corona had a random, Gaussian
magnetic field, anchored to a statistically homogeneous photospheric
field with a known power spectrum $S(k)$.  The density of null points
at a given height, $\rho_N(z)$, scaled with the inverse cube of the
characteristic scale length at that height.  This was expressed in
terms of integrals of the power spectrum, 
$S(k)$, as well as on the mean
of the vertical field, $\bar{B}_z$.   At lower heights it
depended on larger wave numbers (smaller scales).  A perfectly
flat spectrum, such as from ideal white noise or a distribution of
point magnetic charges, introduces no length scale of its own.  In
this case the density decreases with height as
\be
  \rho_N(z) ~=~ {0.040\over z^3} ~~,
\ee
independent of the strength or other properties of the photospheric
field.

% show here that this spectral estimate is relatively accurate for
%   QS.  & that QS spectra are relatively invariant.  one

The spectrum from a magnetogram can be used in the spectral estimate
of \papi\ to compute the distribution of nulls above
observed regions.  Its assumption of randomness and homogeneity are
likely to be fairly valid in quiet Sun regions.  In this work we quantify the
degree to which a potential field extrapolation satisfies the other
assumption, that the field components are Gaussian random variables.
We also test the estimate against the distribution of null points
found directly in an extrapolated field.  We find evidence that the
spectral estimate gives a reasonable approximation of the number of
magnetic null points above a region of quiet Sun, especially above
1,500 km from the magnetogram surface.

The spectra of MDI magnetograms reveal several factors which could
affect the number of null points found either directly or though
spectral estimate.  There is a level of white noise which, while
reduced by averaging, is present in any magnetogram.  We find that
removing its contribution from a spectrum reduces the number of null
points by as much as 85\%.  The defocusing of the telescope, partly on
purpose in the low-resolution mode, steepens the spectrum, causing an
underestimation of the number of null points.  Both of these effects can
be removed from the power spectrum, $S(k)$, yielding a spectral
estimate of null points less affected by instrumental artifacts.

We perform spectral estimates on 562 MDI magnetograms of
quiet Sun during the two solar minima it has observed
(1996--1998 and 2006--2007).  After compensating
for the effects mentioned above, we find the number of coronal null
points above 1.5 Mm to be fairly constant:
$N_n=3.1\times10^{-3}\pm3.0\times10^{-4}\,{\rm Mm}^{-2}$
That is, on average, one such null point
over every $322\,{\rm Mm}^2$ patch of quiet
Sun.  When the Sun is completely free of active regions this amounts to
about $20,000$ null points above the solar surface.

The magnetic spectrum of the quiet Sun is found to be remarkably
consistent over the periods investigated.  The typical spectrum is
consistent with a potential field from magnetic point sources located
about 1.4 
Mm {\em below} the photospheric surface.  The variance and kurtosis of
the magnetograms can be used to infer the size and areal density of
these sources.  Assuming them to have an exponential distribution of
fluxes we find the sources have mean flux, 
$\avg{|\phi|}=1.0\times10^{19}$ Mx and areal 
density $n=0.007\,{\rm Mm}^{-2}$.

The paper is structured as follows.  In the next section we assess the
degree to which potential fields, extrapolated from two particular MDI
magnetograms, satisfy the assumptions of the \papi\
spectral estimate.  We then directly locate all of the null points in
extrapolations from these magnetograms and compare the results to the
spectral estimate.  In \S 3 we apply the spectral 
estimate to 562 MDI magnetograms spanning solar minimum periods in
1996--1998 and 2006--2007.  We find that the number of coronal null
points stays at a relatively constant level, with $\sim10\%$
day-to-day variation.  In \S 4 we show how the generic spectrum of
the quiet Sun photosphere may be interpreted as a random
super-position of identical magnetic elements.  We infer the
properties of the elements and their distribution from the Fourier
spectrum.  Once again, these properties appear relatively constant
over time.  Finally, we discuss what our findings imply about null
points in the actual coronal field above quiet Sun.

\section{Distribution of Nulls}

\subsection{The Gaussianity of the extrapolated field}

\begin{figure}[htp]
\centerline{\psfig{file=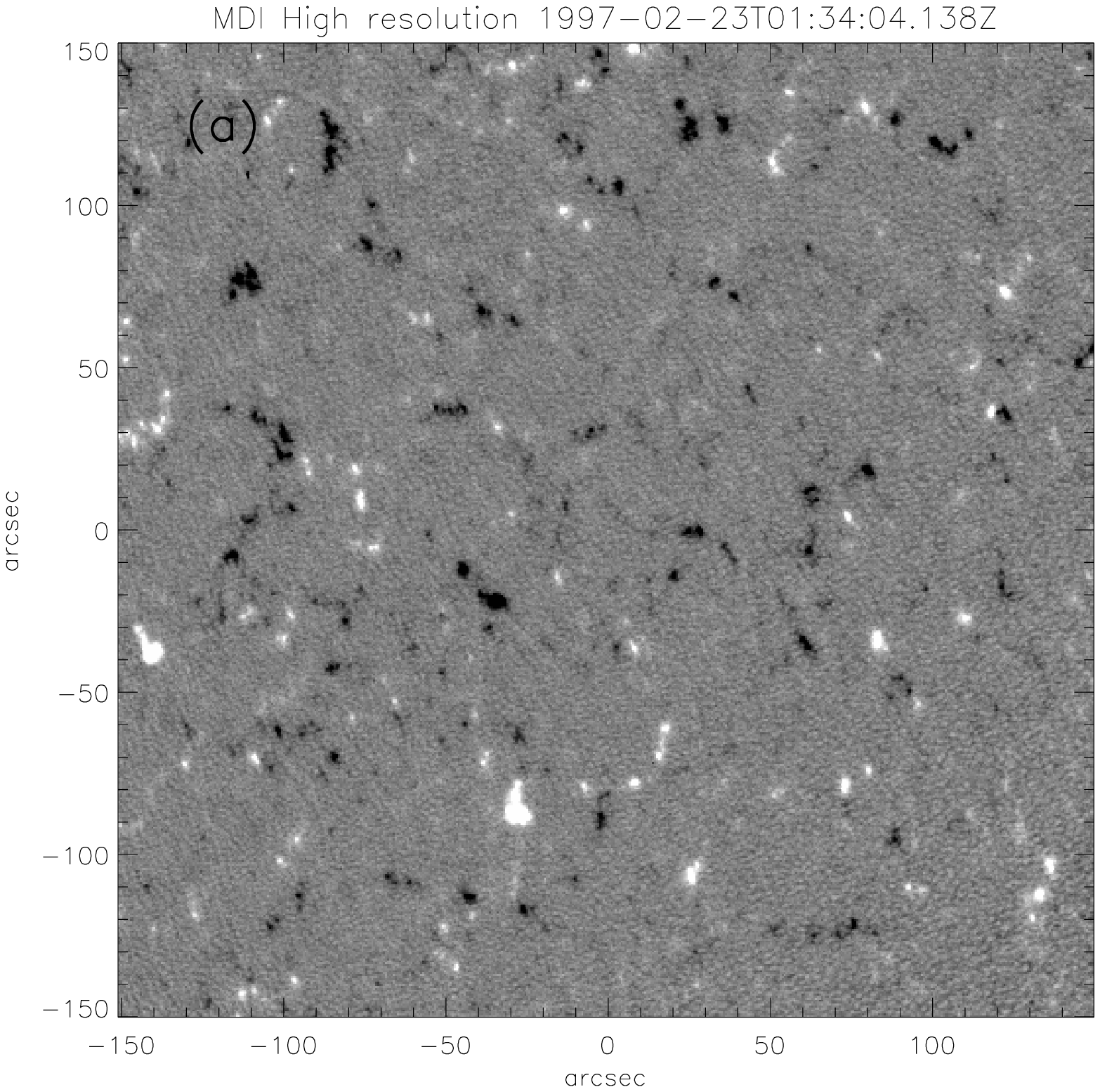,width=2.55in}%
\psfig{file=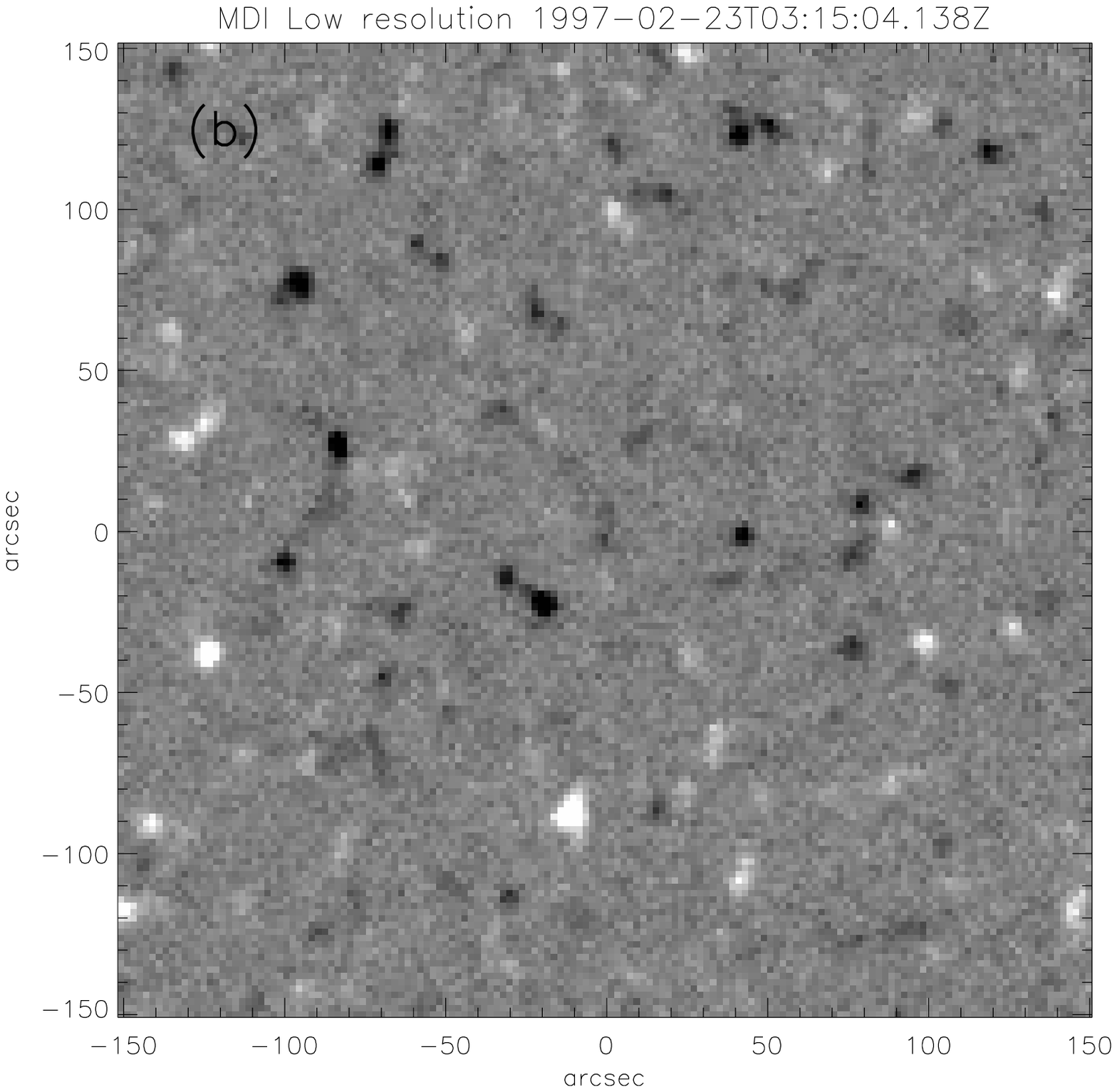,width=2.55in}}
\caption{Magnetograms of a quiet sun area from 23 Feb.\ 1997, made by
averaging five consecutive MDI magnetograms made at one-minute cadence
The grey scale shows the line of sight field between $\pm100$ Gauss.
(a) High resolution magnetograms and (b) Low resolution magnetograms
from 1:45 later.}
	\label{fig:mg}
\end{figure}

The basis for our computation will be magnetograms from regions
of quiet Sun, near disk center.  Figure \ref{fig:mg} shows two examples
of such magnetograms, each made by averaging five one-minute cadence
MDI line-of-sight magnetograms.  
The field of view is $300''\times300''$ centered at disk center.  The
left and right images are from high-resolution and low-resolution data
obtained 1.75 hours apart.

The spectral estimate of \papi\
%\citet[hereafter called \papi]{Longcope2003b} 
yields the spatial density
of null points in a coronal magnetic field satisfying
certain conditions.  It assumes the coronal field to be potential and
to be anchored to a random photospheric field whose statistics are
homogeneous and Gaussian.  These conditions on the photospheric field
(the magnetogram) lead to a coronal field whose components are 
statistically homogeneous and Gaussian.  
From the properties of this field
\papi\ reports the mean density of null points at any height in the
corona.

\begin{figure}[htp]
%\epsscale{1.0}
\psfig{file=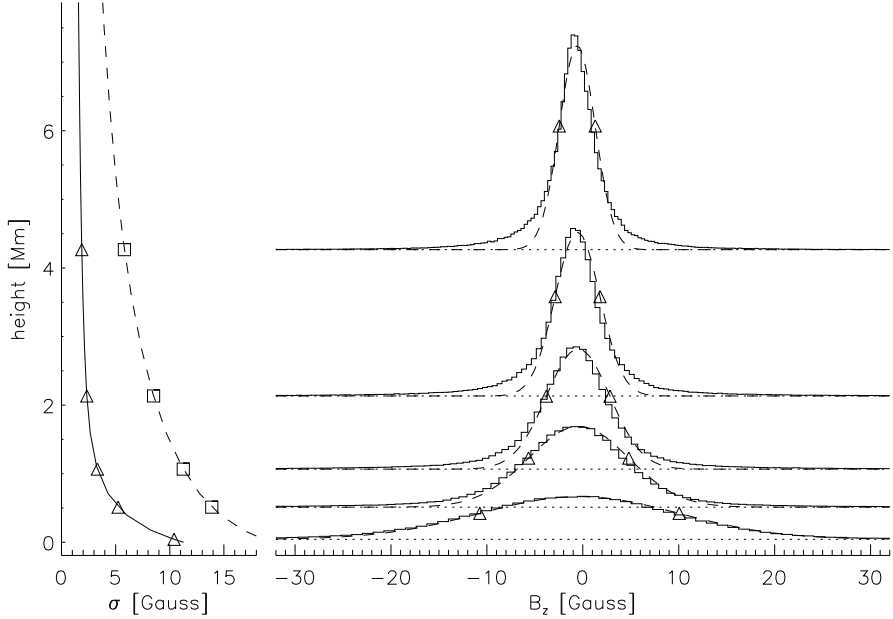,width=5.1in}
\caption{Histograms of the field $B_z$ computed by extrapolating the
photospheric magnetograms to five different heights.  The height, in
Mm, of the slice corresponds to the level of the bottom of the
histogram  on the far left axis.  A Gaussian fit to the central
region is shown as a dashed line, with triangles at
$\pm\sigma_z^{(g)}$.  These values are also plotted against a horizontal
axis on the left.  A solid curve shows the full functional form of
$\sigma_z^{(g)}(z)$.  A dashed line shows $\sigma_z^{(m)}(z)$ on the
same axis (squares occur at heights coincident with the histograms).}
	\label{fig:vgp}
\end{figure}

Extrapolating upward
from the magnetogram of \fig\ \ref{fig:mg}a to a particular height $z$ 
results in a  new field, $B_z(x,y,z)$, whose histogram we find to be 
nearly Gaussian.  Figure
\ref{fig:vgp} shows histograms from several heights as solid lines, and 
the Gaussians best fitting their central portions as dashed curves.
The cumulative distribution from a range of values was fitted, and
the range was iteratively adjusted until it 
contained exactly two standard
deviations to either side.  The standard deviation of this fit,
$\sigma_z^{(g)}$, is indicated by triangles and 
plotted on axes to the left.  The
standard deviation of the entire field, $\sigma_z^{(m)}$, plotted
with a dashed line on the same axis is systematically greater than 
that of the core.

It can be seen in \fig\ \ref{fig:vgp}, that the field $B_z$ has an
approximately Gaussian distribution even well above the photospheric
level ($z=0$).  At the bottom, the Gaussian component has a standard
deviation, $\sigma_z^{(g)}\simeq11.3$ Gauss, comparable to the
intrinsic noise in a five-minute average of MDI magnetograms
\citep{Liu2001}; it decreases with height as the
width of the Gaussian narrows.  The actual
distribution departs from the Gaussian fit principally by an excess
(tail) mostly outside $\pm\sigma_z^{(g)}$.  
Histograms made from the horizontal components of the extrapolated
fields at varying heights look very similar to those of the vertical
field shown in \fig\ \ref{fig:vgp}.

\begin{figure}[htp]
%\epsscale{0.9}
\psfig{file=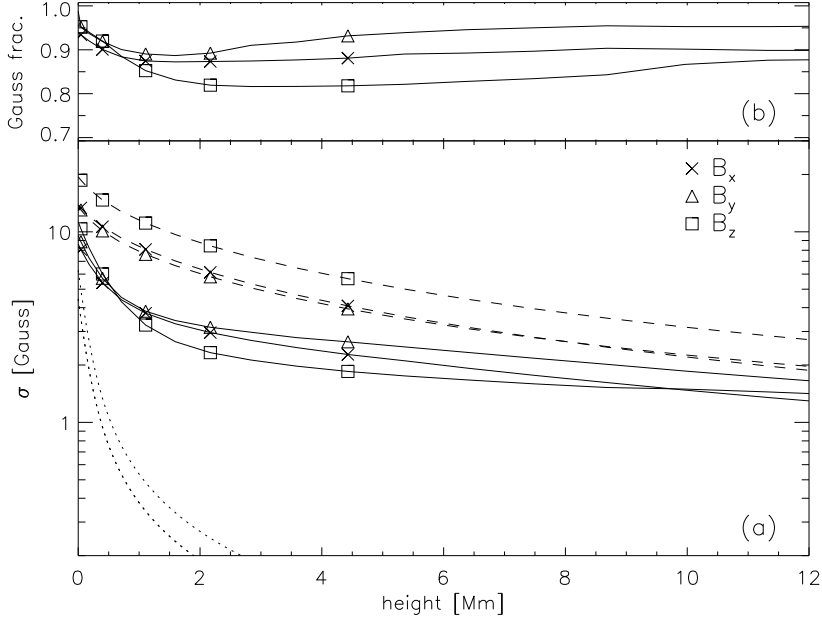,width=5.1in}
\caption{(a) Standard deviations from all three components of the
potential extrapolation of the magnetogram \fig\ \ref{fig:mg}a.  The
solid and dashed curves show $\sigma^{(g)}$ and $\sigma^{(m)}$
respectively for $B_x$ (crosses), $B_y$ (triangles) and $B_z$
(squares), plotted versus height.  The symbols occur at the same
heights as the histograms from \fig\ \ref{fig:vgp}.  Dotted
lines show the standard deviation for vertical 
(upper) and horizontal (lower) field components extrapolated 
from white noise.  
(b) The fraction of each field which is fitted by the Gaussian within
the central 4$\sigma^{(g)}$.}
	\label{fig:sig}
\end{figure}

The Gaussian fit to the central core of each
distribution can be used to quantify the degree to which the actual field
is a Gaussian variable.  The amplitude of the Gaussian component ranges from
$0.97$ at $z=0$ to $0.81$ higher up (see \fig\
\ref{fig:sig}b).  That is to say the field is between $81\%$ and $97\%$
Gaussian.

While a significant component of the magnetic field has a Gaussian
distribution, it is not spatial white noise.
Longcope, Brown and Priest (\papi) \nocite{Longcope2003b} show that a 
potential field extrapolated from a
magnetogram with isotropic power spectral density (psd) $S(k)$ 
has vertical field with variance
\be
  \sigma_z^2(z) ~=~ 2\pi \int_0^{\infty} S(k)e^{-2kz}\, k\, dk 
  ~=~ 2\sigma_x^2(z) ~=~ 2\sigma_y^2(z)~~.
	\label{eq:sig}
\ee
Spatial white noise is uncorrelated from pixel to pixel and therefore has a
flat psd
\be
  S(k)\simeq{\sigma_0^2\over\pi k_c^2}\Theta(k_c-k) ~~,
	\label{eq:Sk_noise}
\ee 
out to the resolution limit $k_c=2\sqrt{\pi}/\Delta x$ for pixel size
$\Delta x$ ($\Theta$ is the Heaviside function).  Using this in
(\ref{eq:sig}) gives the variance due to spatial white noise alone
\be
  \sigma^2_z(z) ~=~ {\sigma_0^2\over8\pi (z/\Delta x)^2}\,
  \left[ 1 - (1+4\sqrt{\pi}z/\Delta x )
  e^{-4\sqrt{\pi}z/\Delta x} \right] ~~.
	\label{eq:sig_noise}
\ee
This functional form, plotted as a dotted line for each component of
the field in \fig\ \ref{fig:sig}, falls off much faster than either
$\sigma^{(g)}$ or $\sigma^{(m)}$.  We therefore conclude that the
Gaussian component consists of contributions with spatial
correlations longer than a single pixel: it is not white noise.

\begin{figure}[htp]
%\epsscale{0.6}
\psfig{file=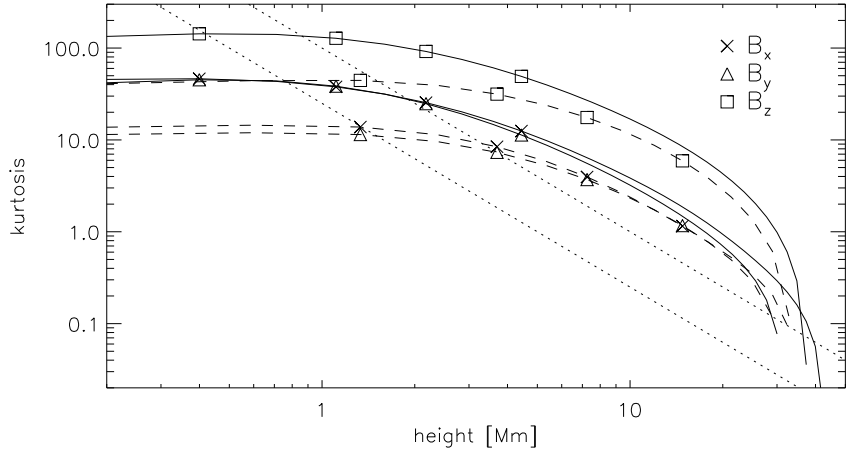,width=4.5in}
\caption{The kurtosis of the fields for $B_x$ (crosses), $B_y$
(triangles) and $B_z$ (squares), plotted versus height.  Solid and
dashed lines are for the high resolution and low resolution
magnetograms respectively.  The dotted lines show the results for
point magnetic sources, taken from \eq\ (\ref{eq:kurt_mpole}).}
	\label{fig:kurt}
\end{figure}

It is common practice to assess the degree to which a variable departs
from Gaussian by its kurtosis statistic, $K$.  The kurtosis is the
fourth moment divided by the square of the variance, less three; 
when the variable is perfectly Gaussian $K=0$.  
Excesses, relative to a Gaussian, in
the tails of the distribution will elevate the fourth moment and lead
to a positive kurtosis.  Figure \ref{fig:kurt} shows that even when
the excess tails account for 3\% of the distribution, they lead to a
large kurtosis.

For comparison, a distribution of $n$ point sources per unit surface
area has a flat power spectrum with no cut-off
($k_c\to\infty$) whose variance is (\papi)
\be
  \sigma_z^2(z) ~=~{n\avg{\phi^2}\over 8\pi z^2} ~~,
	\label{eq:sig_mpole}
\ee
where $\avg{\varphi^2}$ is the variance in the source fluxes.
The kurtosis of this same photospheric field is
\be
  K_z ~=~ {4\over 5\pi}\,{\avg{\varphi^4}\over\avg{\varphi^2}^2}\,
  {1\over nz^2} ~=~ 4K_x ~=~ 4K_y ~~.
	\label{eq:kurt_mpole}
\ee
Both of the expressions above diverge at the photosphere ($z\to0$) as a
consequence of the very strong fields in the neighborhood of a 
point source.  This divergence is one element leading to the failure
of the spectral estimate at low heights, $z<n^{-1/2}$ (\papi).

It is evident from \fig\ \ref{fig:kurt} that the
kurtosis of the actual field is generally large but
remains finite even as $z\to0$.  The large magnitude arises from the
tails in the pdf corresponding to those pixels of the magnetogram in
magnetic elements, where $|B_z|>5\sigma_z(0)\simeq50$ Gauss.  These are the
excess tails of the distribution function, but they are also the
structures which generate field above a few megameters.

\subsection{Assessing the accuracy of the spectral estimate}

A spectral estimate of the magnetic null point density
is found from the isotropic power spectrum,
$S(k)$, of the photospheric magnetogram.  The spectrum determines the
variance, as given by \eq\ (\ref{eq:sig}), and a related quantity
\be
  q^2(z) ~=~ {2\pi \over \sigma_z^2(z)}
  \int_0^{\infty} S(k)e^{-2kz}\, k^3\, dk ~~.
	\label{eq:q2}
\ee
Provided the field has Gaussian statistics the density of null points
in it is found to be
\be
  \rho_N(z) ~=~ G(\bar{B}_z/\sigma_z)\, q^3(z) ~~,
	\label{eq:rho}
\ee
where $G(\zeta)$ is a dimensionless function defined in \papi.
The function depends on the mean vertical field strength $\bar{B}_z$ and also
weakly on another moment of the spectrum.
Provided the polarities are sufficiently balanced that
$\bar{B}_z\ll\sigma_z$ over the relevant range of heights, we find
$G\sim 2\times10^{-2}$ for most spectra of quiet Sun magnetograms.

\begin{figure}[htp]
%\epsscale{1.0}
\psfig{file=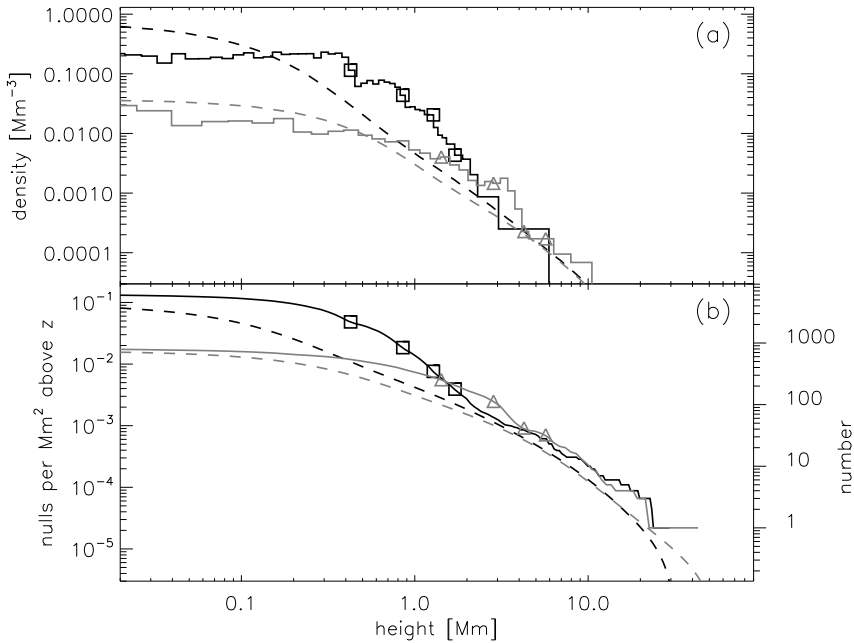,width=5.1in}
\caption{The density (a) and column density (b)
of null points in potential fields
extrapolated from the MDI magnetograms in \fig\ \ref{fig:mg}.  Dark and
light curves correspond to the high resolution and low resolution
data respectively.  The solid curves are computed by finding
the nulls on a gridded extrapolation; the dashed curves are the
spectral estimates.  Symbols appear at heights of one, two, three and
four horizontal pixel sizes ($z=\Delta x$, 
$2\Delta x$, $3\Delta x$ and $4\Delta x$).
For panel (b) the left axis reports the column density in nulls per
Mm$^2$ above a given height, while the right axis gives the total number 
over the entire magnetogram.}
	\label{fig:nd}
\end{figure}

We assess the accuracy of the approximation offered by the spectral
estimate by direct comparison.  Both magnetograms from \fig\
\ref{fig:mg} are extrapolated onto a rectilinear grid with periodic
lateral boundaries.  Between the grid points the magnetic field is
defined using tri-linear interpolation.  A magnetic null point is the
point where all three components of this interpolated field vanish
together.  These null points are located using the algorithm of
Haynes and Parnell (2007).
\nocite{Haynes2007}  The number of null points found above a given
height $z$ is divided by the photospheric area to yield a null-point
column density, $N_n(z)$. The right and left axes of \fig\ \ref{fig:nd}b
shows the number and column densities found this way, respectively.

There are more null points in the high resolution case (dark curves)
than in the low resolution (6135 {\em vs.} 664).
The majority of the excess null points lie in a layer about one Mm
thick.  Over heights above one Mm the predicted null columns (dashed
curves on \fig\ \ref{fig:nd}b) from the high resolution and low
resolution cases are quite similar.  The greatest disagreement between
the actual null column, $N_n(z)$, and the spectral estimate occurs
within this same one Mm bottom layer.  Above this height the two estimates 
are within a factor of two of reality for both resolutions.  Since the
bottom layer of the high resolution field has the largest kurtosis it
is perhaps unsurprising that the greatest 
disagreement with the spectral estimate occurs there.  
Perhaps more surprising is
that kurtosis as large as 40 (the maximum of the low resolution data)
leads to disagreement by not more than a factor of two.

\section{The quiet Sun null column in general}

\subsection{Comparing nulls from different magnetograms}

\label{sec:psd}

\begin{figure}[htp]
%\epsscale{1.0}
\psfig{file=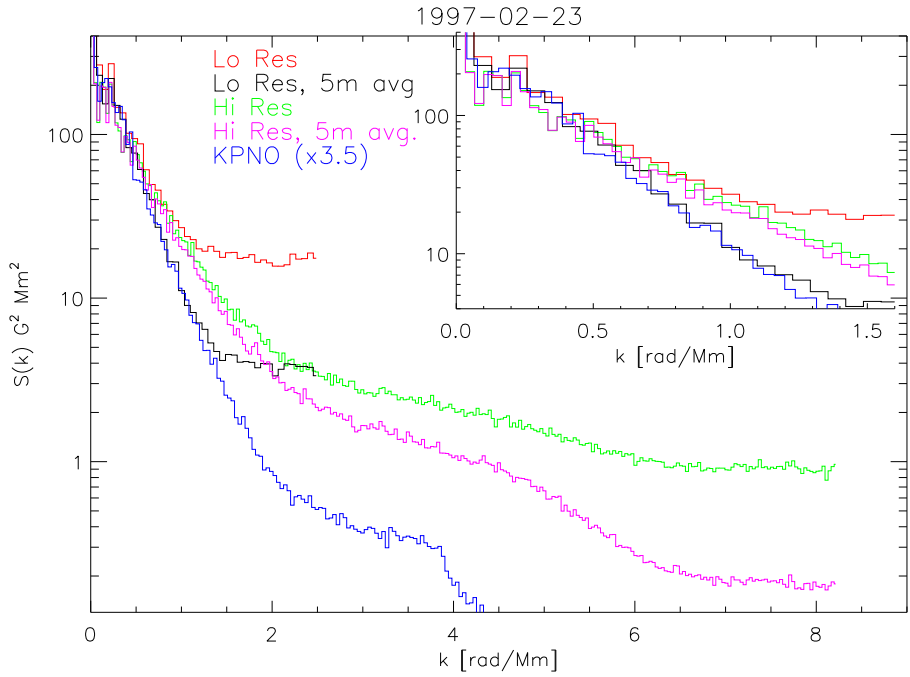,width=5.1in}
\caption{The isotropized spectra for $300''\times300''$ magnetograms
of quiet Sun centered at disk center on 23 Feb.\ 1997.  Curves of
different colors are $S(k)$ for 30-second MDI exposures at high (green)
and low (red) resolution; 5-minute averages at high (magenta) and
low (black) resolution; and Kitt Peak magnetogram (blue).  The inset
is an expansion of the region at low wavenumbers.}
	\label{fig:spec}
\end{figure}

The extent to which, for levels above one Mm, the spectral estimates
of high resolution and low resolution agree can be understood
from a comparison of their power spectra $S(k)$, as shown in
\fig\ \ref{fig:spec}.  A fast Fourier transform (FFT) of the
$L_x\times L_y$ magnetogram returns an array of complex 
coefficients, $\hat{B}_{m,n}$, for the expansion
\be
  B_z(x,y) ~=~ \sum_{m,n}\hat{B}_{m,n}\,
  \exp[ \,i\,m\Delta k_x x + i\,n\Delta k_y y \,] ~~,
\ee
where $\Delta k_x=2\pi/L_x$ and $\Delta k_y=2\pi/L_y$.  The 
quiet-Sun magnetograms have very few features at their edges which
might suffer discontinuity under periodic repetition so we find it
unnecessary to employ any windowing 
function \citep{Press1986}.

The two-dimensional power spectrum at
wave number $\kvec=m\Delta k_x\xhat+n\Delta k_y\yhat$ is
\be
  S(\kvec) ~=~ {|\hat{B}_{m,n}|^2\over \Delta k_x\Delta k_y} ~~.
	\label{eq:Sk}
\ee
This function is defined only on points of the Fourier-space grid, but
approximates a continuous function.  The denominator in
\eq\ (\ref{eq:Sk}) eliminates from $S(\kvec)$ any dependance on pixel
size or field of view.

The isotropized power spectrum, $S(k)$, is computed by averaging all values
of $S(\kvec)$ within an annulus of Fourier-space.
We use annuli of radii extending to a
maximum wave number $k_c=2\sqrt{\pi}/\Delta x$,
chosen so that the total Fourier-space area, $\pi k_c^2$, matches
that occupied by the actual magnetogram.  
The cut-offs for high and
low resolution MDI magnetograms are therefore 
$8.2$ rad/Mm and $2.5$ rad/Mm respectively since these have
pixel sizes of $\Delta x=0.43$ Mm and $\Delta x=1.42$ Mm in 
this particular magnetogram.

The spectrum in \eq\
(\ref{eq:Sk}) is normalized so that, 
\[
  \avg{B_z^2} ~=~ \sum_{m,n}|\hat{B}_{m,n}|^2\, ~\simeq~
  \int S(\kvec)\, d^2k ~=~2\pi\int_0^{k_c}S(k)\, k\, dk ~~,
\]
where the angled brackets are a spatial average.
The variance is found from the same expression but
excluding from the sum $(m,n)=(0,0)$.  The variance at height $z$
above the magnetogram plane, \eq\ (\ref{eq:sig}), includes an
exponential factor from the potential field extrapolation.

Each spectra from an MDI magnetogram in \fig\ \ref{fig:spec} has
a range of white noise over its highest wave numbers.
We estimate its level, $S_n$, from the median value 
of $S(k)$ over the highest 15 spectral
bins.  The noise level is reduced by exactly $1/5$ in the
averages strongly suggesting an origin in Poisson counting
noise, uncorrelated between frames \citep{Liu2001,DeForest2007}.
The flat noise floors contribute 
$\sigma^2_z(0)=\pi k_c S^2_n = (8.8\,{\rm G})^2$ and
$(6.3\,{\rm G})^2$ to the low and high-resolution
variances respectively.  We noise-correct each spectrum by 
subtracting from the entire curve 
the flat noise floor, $S_n$, determined from the top 15
spectral bins.

\begin{figure}[htp]
%\epsscale{1.0}
\psfig{file=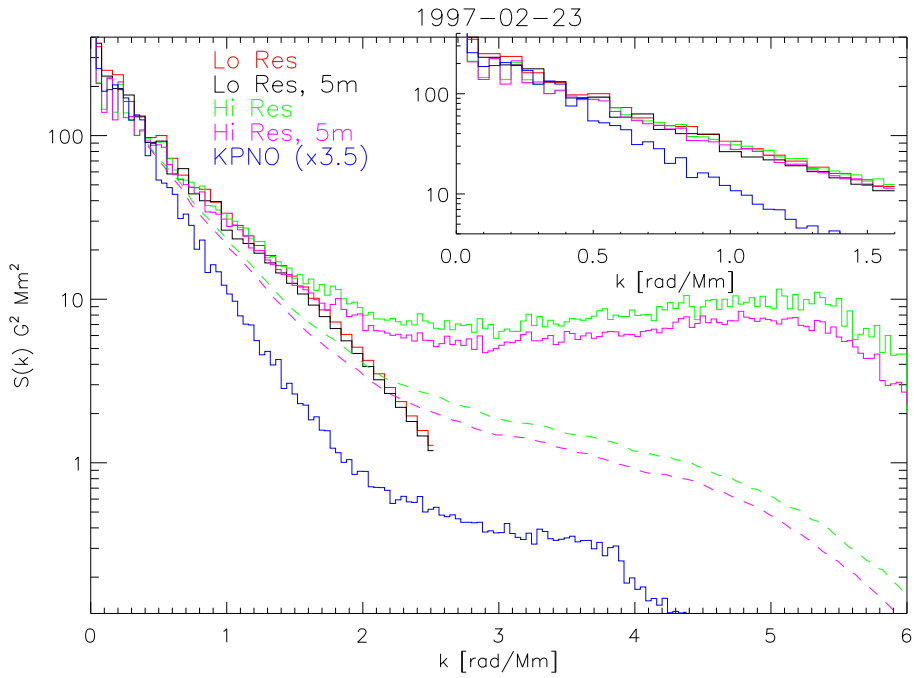,width=5.1in}
\caption{The same spectra from \fig\ \ref{fig:spec} after removal of
white noise and correction for the MTF.  No corrections are
performed on the Kitt Peak spectra (blue) which is retained for
reference.  Dashed curves show the high resolution spectra 
before it was corrected for the MTF, but after white noise was removed.}
   \label{fig:spec_mtf}
\end{figure}

Each of the noise-corrected spectra appears to have been 
steepened by
the modulation transfer function (MTF), $M(k)$,
of its imaging system \citep{Born1980}, and must be corrected to
$S(k)/M^2(k)$.  The most conservative correction to the
high-resolution spectrum is to assume it to be diffraction-limited and
take $M(k)=M_d(k)$, given in an appendix.  MDI is intentionally
defocussed in its low-resolution mode, leading to an MTF we
approximate as 
$M^2(k) = M_d^2(k)/( 1 + \ell_0^4k^4 )$ where $\ell_0\simeq1.6''$ at
this time (see appendix).  Performing these corrections restores all
four MDI spectra to similar forms, over low wave-numbers,
as shown in \fig\ \ref{fig:spec_mtf}.
The Kitt Peak spectrum (blue) remains
below all others since it has not been divided by its MTF, part of
which presumably arises from atmospheric seeing
\citep{Hufnagel1964,Abramenko2001}.

There is, over the range $k>1.5\,{\rm rad/Mm}$, an excess in
high-resolution power, relative to low-resolution; this excess is
further amplified by the MTF correction.  The excess might 
be partly an artifact from the granular motions
during the brief intervals between different filter-gram exposures
\citep{DeForest2007}.  Its origin, whatever that may be, is evidently 
correlated several-fold longer than five minutes, 
since averaging diminishes it by less than a factor of five.
We next show that this power does not significantly affect
the null densities at heights where potential field extrapolation is
warranted.

\begin{figure}[htp]
%\epsscale{1.0}
\psfig{file=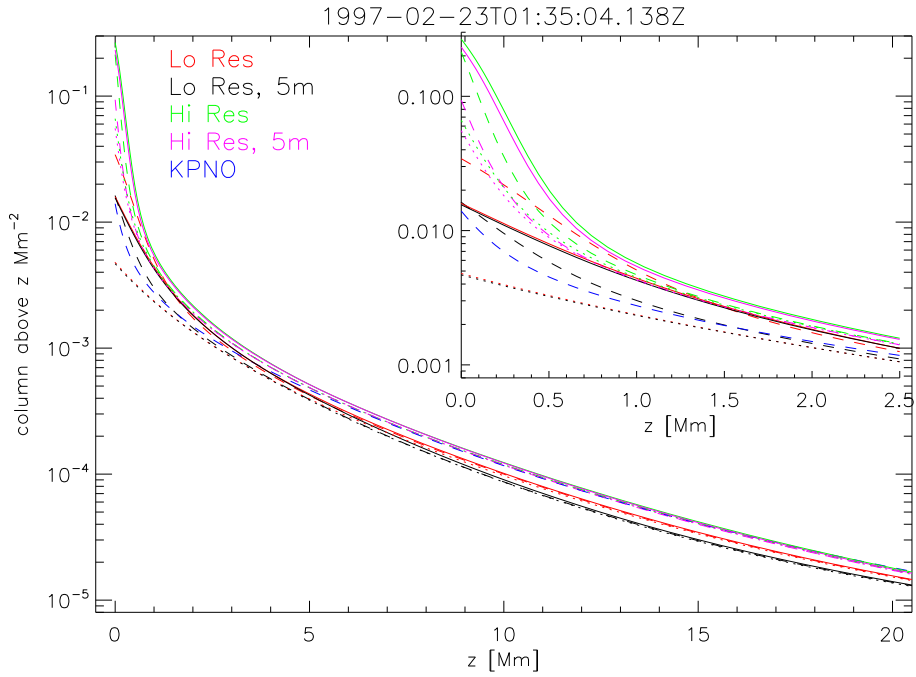,width=5.1in}
\caption{Spectral estimates of the null column densities, $N_n(z)$,
computed from each of the spectra in \figs\ \ref{fig:spec} 
and \ref{fig:spec_mtf} using the same color
code.  Dashed curves are from
the original spectra (\fig\ \ref{fig:spec_mtf});
solid curves are from spectra
from which white noise has been subtracted and which 
have been divided by the MTF (\fig\ \ref{fig:spec_mtf}).  
Dotted lines show the curves after noise subtraction without MTF
correction.}
	\label{fig:nd2-23}
\end{figure}

The spectra from \fig\ \ref{fig:spec} are used in the Laplace
transform integrals, (\ref{eq:sig}) and (\ref{eq:q2}), from which are
computed the null density $\rho_N(z)$, according to \eq\ (\ref{eq:rho}).  
This density is integrated downward from some
upper height yielding the null column density, $N_n(z)$, plotted in
\fig\ \ref{fig:nd2-23}.  The exponential factor, $e^{-2kz}$, in each
of the spectral integrals means that the null density at a given
height $z$ depends principally on the spectrum over wave numbers
$k<1/z$.  Thus heights above $z=1$ Mm depend mostly on the spectrum
below (to the left of) $k=1\,{\rm rad/Mm}$.  As discussed above, and
evident in \fig\ \ref{fig:spec_mtf}, all MDI spectra are fairly similar 
over this range.  It is therefore natural that all five solid 
curves in \fig\ \ref{fig:nd2-23} follow similar paths above one Mm.

Noise contributes significant power to the highest wavenumbers,
creating a population of artificial null
points in an extrapolation.
The null column densities after subtraction of this noise,
the dotted curves in \fig\
\ref{fig:nd2-23}, lie below the corresponding uncorrected
(dashed) curves.  The fraction of nulls attributable to noise alone is
lower in five-minute averages than one-minute magnetograms, since the
former have lower noise.  The effect is most pronounced in one-minute
low-resolution magnetograms, for which $85\%$ of all nulls are due to
noise.

The small-scale structure revealed in high resolution magnetograms
translates into complex fields with high null densities at the
lowest level of the potential magnetic field.  While potential fields are
frequently used to approximate the quiet Sun corona, they are of
questionable validity in the 
lowest layers, where the plasma density and pressure are liable to
play their most significant role.  Above a height of 1,500 km (1.5
Mm) the potential field 
approximation may be on better footing.  It appears from \fig\
\ref{fig:nd2-23} that most of the estimates based on potential field
extrapolation, are in reasonable agreement above this particular
height (at $z=1$ Mm, for example, the high and low resolution
estimates appear to disagree to a greater extent).  For these reasons
we hereafter focus on estimating the total column of null points above
$z=1.5$ Mm, denoted $N_n(1.5\,{\rm Mm})$.

\subsection{The generic behavior}

The foregoing sections have considered different magnetograms of the
same region of quiet Sun (23 Feb.\ 1997).  In order to generalize our
conclusions we perform similar analysis on magnetograms spanning solar
minimum from 2006 and 2007.  The same central $300''\times300''$ sub-image
is extracted from one full-disk (low resolution) 
daily MDI magnetogram every second day.  
It is analyzed provided there is no obvious active region
field, or preponderance of one polarity, and provided there is less
than $8\times 10^{19}$ Mx in pixels above 200 G.  This value was
selected after some trial and error to eliminate plage regions whose
presence would violate our assumption of homogeneity.  Separate
analyses are performed for one-minute daily magnetograms 
and for those composed of 5-minute averages.

\begin{figure}[htp]
%\epsscale{0.9}
\psfig{file=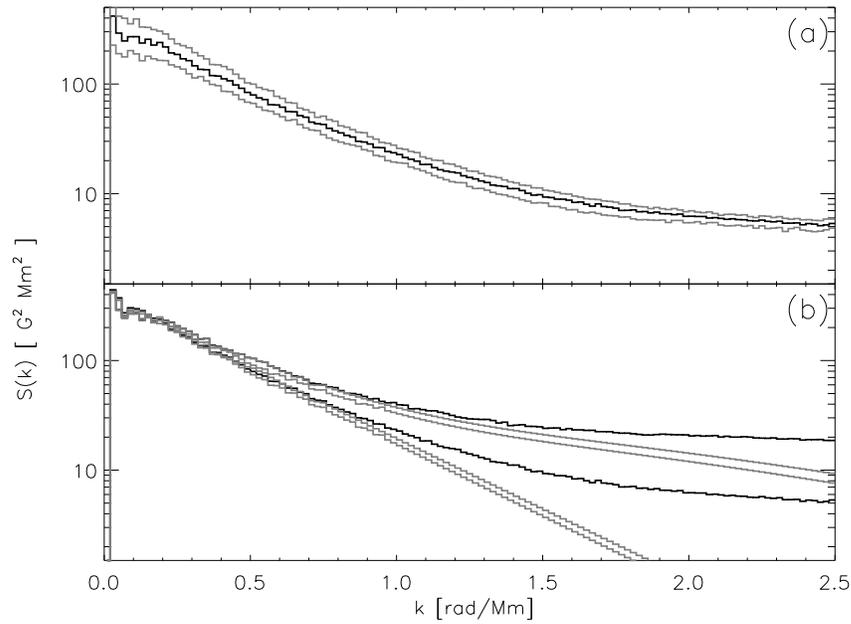,width=5.1in}
\caption{Median spectra from magnetograms in 2006 and 2007.
(a) The median spectrum
(dark) and the upper and lower quartiles (light curves) from the
five-minute averages.  (b) The median from 118
one-minute magnetograms (upper curve) and 114 
5-minute averages (lower curve).  The lighter curves show the results
of subtracting white noise (lower pair) and then dividing by the MTF (upper
pair).}
	\label{fig:med_spec}
\end{figure}

The isotropized spectra from all quiet Sun areas turn out to be very
similar to one another, closely resembling the example from the previous
section.  Figure \ref{fig:med_spec} shows a median spectrum: the
the median value, at each wave-number, from over 110 
quiet Sun spectra of each low-resolution kind.
The upper and lower quartile points, bracketing half of all values, 
are plotted as
lighter curves.  This clearly shows that the most significant variation
occurs at wavenumbers below 0.2 rad/Mm (i.e.\ wavelengths longer
than 30 Mm).

Removing the white noise floor by first
subtracting it and then extrapolating the noisier portions
yields the lower light spectra in \fig\
\ref{fig:med_spec}b.  This is then divided by the empirically 
determined MTF, $M^2(k)=M_d^2(k)/(1+\ell_0^4k^4)$, where $\ell_0=1.04''$ as
described in the appendix.  The results are the two upper light
curves.

\begin{figure}[htp]
\psfig{file=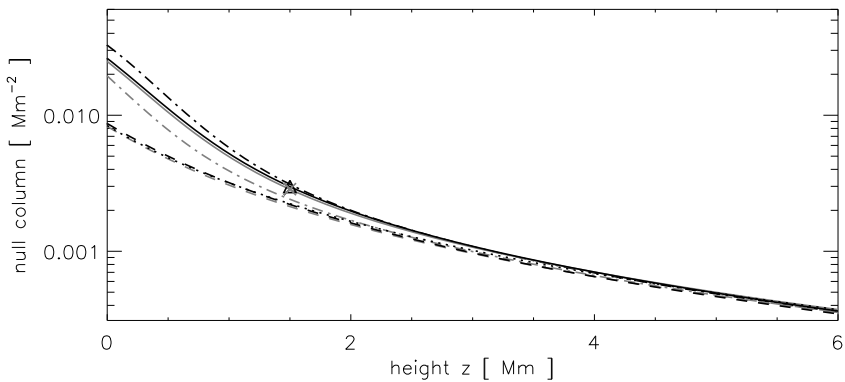,width=5.1in}
\caption{The column densities from each of the median spectra in \fig\
\ref{fig:med_spec}b.  Dark and light curves are for one-minute magnetograms
and five-minute averages respectively.  Solid curves are from the
entire spectra and dashed curves are from the noise-corrected
spectra.  A dotted curve is an empirical fit.}.
	\label{fig:med_den}
\end{figure}

The column density found from these various versions of the median
spectra are shown in \fig\ \ref{fig:med_den}.  The solid curves of
each shade are the result of removing noise and correcting for 
the MTF.  The broken
curves above these are from the raw spectra, while the dashed 
curves below are after noise has been removed 
but no MTF correction performed.
Symbols mark the values of these at $z=1.5$ Mm:
$3.0\times10^{-3}$ Mm$^{-2}$ and $2.9\times10^{-3}$
Mm$^{-2}$ respectively.

The median spectra are clearly very close to pure exponentials,
$S(k)\sim e^{-2kd}$.  This functional form suggests the empirical model
for the null column density of
one-minute magnetograms (dotted line)
\be
  N_n(z) ~\simeq~ {0.021\over (z+d)^2} ~~,
	\label{eq:LBP_col}
\ee
where the numerator and $d=1.6$ Mm are found from fitting.

\begin{figure}[htp]
%\epsscale{1.0}
\psfig{file=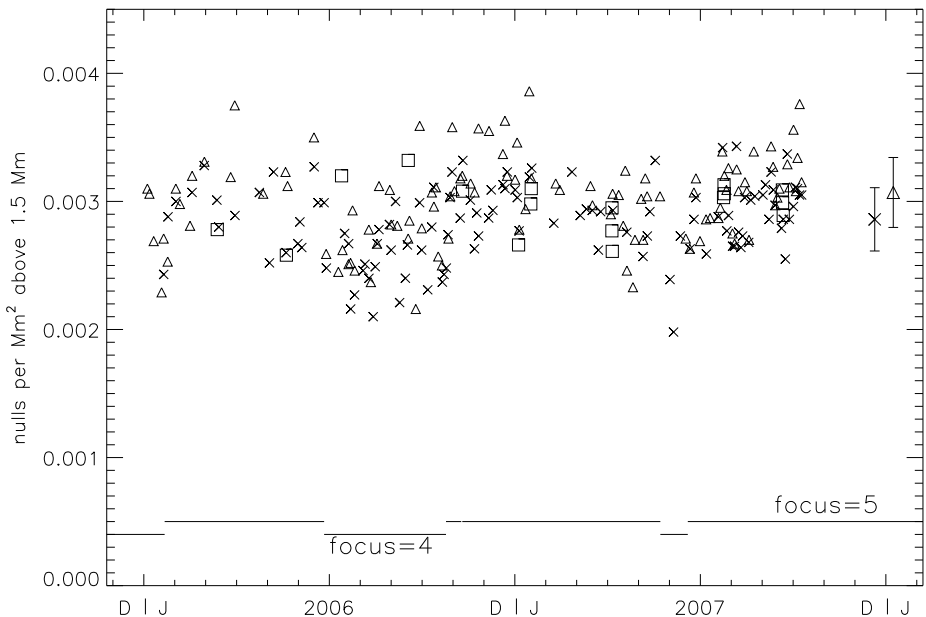,width=5.1in}
\caption{The column of null points above $z=1.5$ Mm found from quiet
sun magnetograms spanning 2 years of solar minimum.  
Triangles and crosses denote
one-minute and five-minute low-resolution magnetograms respectively;
squares are from high-resolution magnetograms.  Error
bars on the right show one standard deviation of scatter about a
90-day running mean.  Solid lines along the bottom show the reported
focus position of the MDI telescope.}
	\label{fig:smry}
\end{figure}

The null column above a given height does vary between magnetograms, as
shown in \fig\ \ref{fig:smry}.  There appear to be 
random fluctuations superimposed on a much slower variation.  We
estimated the
dispersion about the trend, $2.7\times10^{-4}\,{\rm Mm}^{-2}$
for the one-minute magnetograms and $2.5\times10^{-4}\,{\rm Mm}^{-2}$ for
five-minute averages, by subtracting a 90-day running mean.  
Considering the entire data set of one-minute
magnetograms, including its slow trend, we find the column of null
points above $z=1.5$ Mm to be
\[
  N_n(1.5\, {\rm Mm}) ~=~ (3.1\pm 0.3)\times10^{-3}\, {\rm Mm}^{-2} ~~.
\]
(The errors quoted above represent rms variation; the error in the mean is far smaller.)
The value from five-minute magnetograms is slightly
lower: $(2.9\pm 0.3)\times10^{-3}$.  We believe this value is lower 
partly because of motional blurring during the averaging interval, 
so we adopt the 
one-minute-magnetogram value as the more accurate of the two.

We believe the slow variation in $N_n(1.5)$ is due to gradual changes
in the MTF of the imaging system.  
The reported focus position of the
instrument is shown as a solid curve long the bottom.  The focus is
periodically changed to compensate for changes in the instrument
itself.  Our
estimate of the MTF was performed through cross-comparison to the high
resolution images, plotted as squares, all 
while the instrument was in focus position 5.  This is the MTF by
which we are correcting the spectrum, so we believe that the value in
the middle of this interval is probably the most accurate.  
The MTF correction raises the low resolution columns by about 40\%,
thereby making them more consistent with high resolution values.
Omitting that step gives null columns which could be considered lower
bounds on the actual values: medians are
$N_n(1.5)=2.2\times10^{-3}\,{\rm Mm}^{-2}$ 
for one-minute 
and $2.1\times10^{-3}\,{\rm Mm}^{-2}$ for five-minute.

\begin{figure}[htp]
%\epsscale{1.0}
\psfig{file=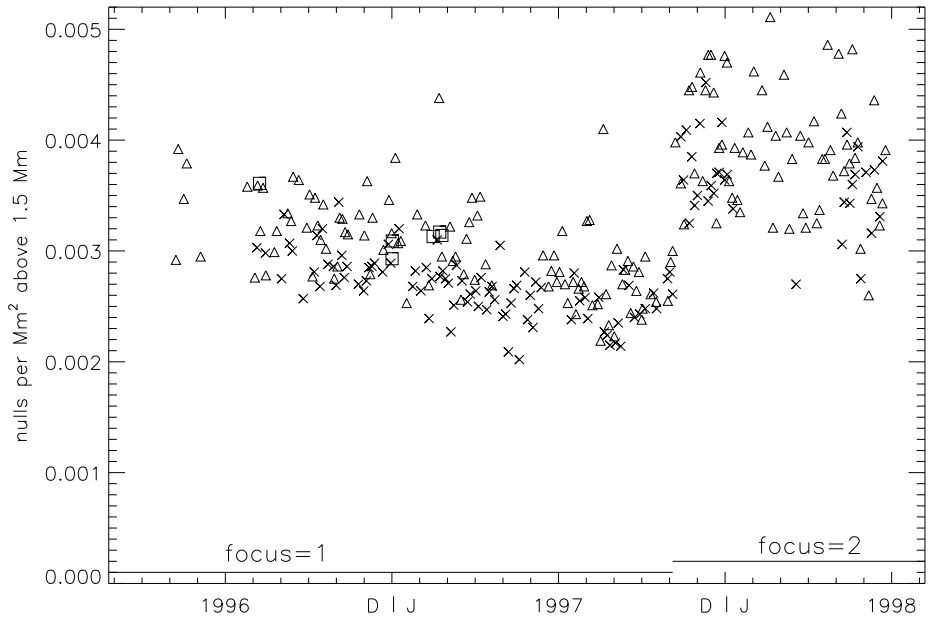,width=5.1in}
\caption{Null columns during the solar minimum from 1996-1998.  Plot
format is the same as in \fig\ \ref{fig:smry}.}
	\label{fig:smry96}
\end{figure}

A comparable plot from the previous solar minimum, \fig\
\ref{fig:smry96}, shows more pronounced evolution on long time scales.  This
data comes from the very beginning of the SOHO mission, when the
instrument was probably changing most rapidly.  There is a very 
large change in
Nov. 1997 coincident with the refocussing of the telescope.  The
later refocussing events in \fig\ \ref{fig:smry} produce much less
obvious changes, either because they are more frequent or because
there is less dramatic change between them.  Due to the
larger, and more rapid change we did not attempt an estimate of the
MTF.  Instead we used the empirical form from the later phase, with
$\ell_0=1.6''$ found from fitting the 23 Feb.\ 1997 data described
above.  This restores the typical values to the same range as they
were in the 2006/2007 data set.

\section{Modeling the quiet Sun spectrum}

\subsection{A synthetic magnetogram}

Some insight into the origins of the quiet sun magnetic field may be
gleaned from its spectrum.  Consider a model photospheric 
field composed of distinct elements, each characterized by the same
normalized shape function, $f(\xvec)$.  Elements at locations,
$\xvec_j=(x_j,y_j)$, 
with fluxes, $\phi_j$, will create a vertical field
\be
  B_z(\xvec) ~=~\sum_j\phi_j\,f(\xvec-\xvec_j) ~~,
	\label{eq:composite}
\ee
in the photospheric plane.
For simplicity let us assume that the fluxes are random, 
have zero mean, and are uncorrelated:
$\avg{\phi_i\phi_j}=\avg{\phi^2}\delta_{ij}$.  We also assume that the
fluxes are independent of their locations $\xvec_j$ which are distributed
randomly with areal density $n$.   Under these assumptions the power
spectrum of the photospheric field is
\be
  S(\kvec) ~=~ n\avg{\phi^2}\,|\hat{f}(\kvec)|^2 ~~,
\ee
where $\hat{f}(\kvec)$ is the Fourier transform of the shape function.

A purely exponential spectrum can be interpreted as being
composed of individual flux elements each with an exponential Fourier
transform such as
\be
  f(\xvec) ~=~ {1\over2\pi}{d\over (|\xvec|^2 + d^2)^{3/2}} ~=~
  \int {e^{-|\kvec|d}\over(2\pi)^2} \, e^{i\kvec\cdot\xvec}d^2 k ~~.
	\label{eq:shape}
\ee
This function also happens to be the vertical potential
field due to a unit point charge located at $z=-d$; it is the
potential field Green's function.

\begin{figure}[htp]
\psfig{file=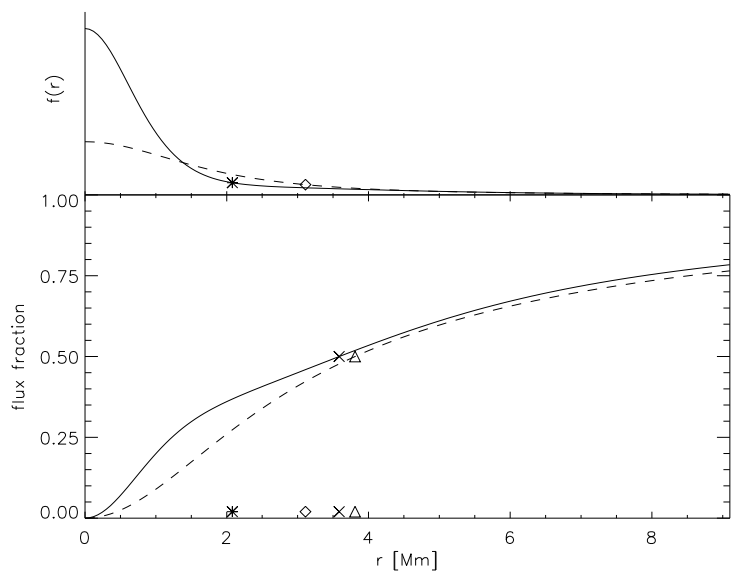,width=4.5in}
\caption{The structure of a magnetic element inferred from the power
spectrum of a magnetogram.  The top curves show the shape functions,
$f(r)$, from \eq\ (\ref{eq:shape}), for an uncorrected
exponential spectrum with $d=2.2$ Mm (dashed) and for the shape after
correction for the low-resolution MTF with $\ell_0=1.6''$ (solid).
Symbols on each curve mark $r_*$.  The vertical scale is the
same for each curve and makes $2\pi\int f(r)\,r\,dr=1$.
The bottom panel shows the total fraction of flux with a given radius,
and symbols mark the radius of half flux.}
	\label{fig:shape}
\end{figure}

The observed spectrum, before MTF-correction, is very close to an
exponential,  $S(k)\simeq e^{-2kd}$.  The exponent is thus directly
related to a ``depth'' at which point sources are submerged.  On the
other hand, the spectrum after correction,
$S(k)\simeq e^{-2kd}/M^2(k)$, has an inverse Fourier transform 
which is not a simple function, and does not admit such a suggestive
interpretation.  Figure \ref{fig:shape} shows the functions $f(r)$
for both the uncorrected (dashed) and MTF-corrected (solid) spectra
from 23 Feb.\ 1997.  The integrated field (bottom) shows that for
each case half of all flux is confined to a similar $r_0\simeq4$
Mm circle.

The variance of a photospheric field composed of random elements is
\be
  \sigma_p^2 ~=~ n\avg{\phi^2}\int f^2\, d^2x ~=~
  {n\avg{\phi^2} \over 4\pi r_*^2} ~~,
	\label{eq:sig_subpole}
\ee
where $r_*$ characterizes the radius of the shape function, $f(r)$, 
through the integral of its square.  The parameter is defined to match
the root-mean-squared radius of a Gaussian shape.  The uncorrected
shape function, \eq\  (\ref{eq:shape}), yields, $r_*=\sqrt{2}d$, whose
introduction into \eq\ (\ref{eq:sig_subpole}) returns the variance of
the vertical field at a height $d$ 
above a distribution of point charges, i.e.\ \eq\
(\ref{eq:sig_mpole}).  The MTF-corrected shape function is
notably narrower (solid line in \fig\ \ref{fig:shape}), under the
anticipation that it is broadened by the point spread function to
produce the broader observed shape (dashed) with its larger radius
$r_*$.

The kurtosis of the composite field given by expression
(\ref{eq:composite}) is
\be
  K_p ~=~ {1\over n}\,{\avg{\phi^4}\over \avg{\phi^2}^2}
  \,{\chi\over 4\pi r_*^2} ~~,~~ \chi ~=~ (4\pi r_*^2)^3
  \int f^4\, d^2x ~~,
	\label{eq:kurt_subpole}
\ee
where the dimensionless coefficient $\chi$ depends only on the shape
of the individual elements, $f(r)$.  It is unity for a perfectly flat
function and greater as the shape becomes more peaked.  For
charges submerged at a depth $d$, the value $\chi=32/5$ can be
inferred by comparison between expressions (\ref{eq:kurt_mpole}) and
(\ref{eq:kurt_subpole}) and use of the relation 
$r_*=\sqrt{2}d=\sqrt{2}z$.

The dimensionless ratio,
$\xi=\avg{\phi^4}/\avg{\phi^2}^2$, depends only on the distribution of
element amplitudes, and not on their size or shape.  Flux amplitudes
distributed uniformly
or exponentially will be characterized by 
$\xi=9/5$ or $\xi=6$ respectively.  Parnell (2002) \nocite{Parnell2002} 
found that
fluxes of each polarity were distributed with a hybrid between an
exponential and power-law known as a Weibull distribution.  The
particular Weibull distributions reported are characterized by
values of $\xi$ between $22$ and $30$.  These are larger than for
purely exponential distribution since broader wings lead to more common
occurrence of large fluxes.

The variance of Gaussian white noise, $\sigma_n^2$, adds to expression
(\ref{eq:sig_subpole}) to give the total variance in a magnetogram,
$\sigma_0^2=\sigma_p^2+\sigma_n^2$.  The 
Gaussian noise has zero kurtosis, so its addition will decrease the
kurtosis of the composite, \eq\ (\ref{eq:kurt_subpole}), 
by the factor $\sigma_p^4/\sigma_0^4$.  
The areal density of flux elements follows directly from the
magnetogram's kurtosis, $K_0$, and properties of the elements
\be
  n ~=~ \left(1 - {\sigma_n^2\over\sigma_0^2}\right)^2
  {\chi\over 4\pi r_*^2}\,{\xi\over K_0} ~~.
	\label{eq:n_from_K}
\ee
Having found the density, expression (\ref{eq:sig_subpole}) can be
used to find the root-mean-square (rms) amplitude of the flux,
$\avg{\phi^2}^{1/2}$.

Using the above reasoning we may model a quiet sun magnetogram as a
combination of Gaussian white noise and a random distribution of flux
elements.  The magnetogram in
\fig\ \ref{fig:mg}b, for example, has a power spectrum which,
after removal of white noise, 
fits an exponential $S(k)\sim e^{-2kd}$, with
$d\simeq 2.2$ Mm.  The magnetogram has a standard deviation
$\sigma_0=15.8$ G, only $\sigma_n=8.8$ G of which comes from 
white noise.  Assuming an exponential
distribution of flux amplitudes ($\xi=6$) the kurtosis, $K_0=38$, implies a
flux element density, 
$n=0.0039\,{\rm Mm}^{-2}$, from \eq\ (\ref{eq:n_from_K}).  
Equation (\ref{eq:sig_subpole}) then
gives the rms flux, $\avg{\phi^2}^{1/2}=2.3\times 10^{19}$ Mx.
The exponential flux distribution and zero mean value can be used 
to convert the rms to $\avg{|\phi|}=1.6\times10^{19}$ Mx. 
% slightly larger than values found by
% previous investigators \citep{Schrijver1997b,Parnell2002}.  

\begin{figure}[htp]
\centerline{\psfig{file=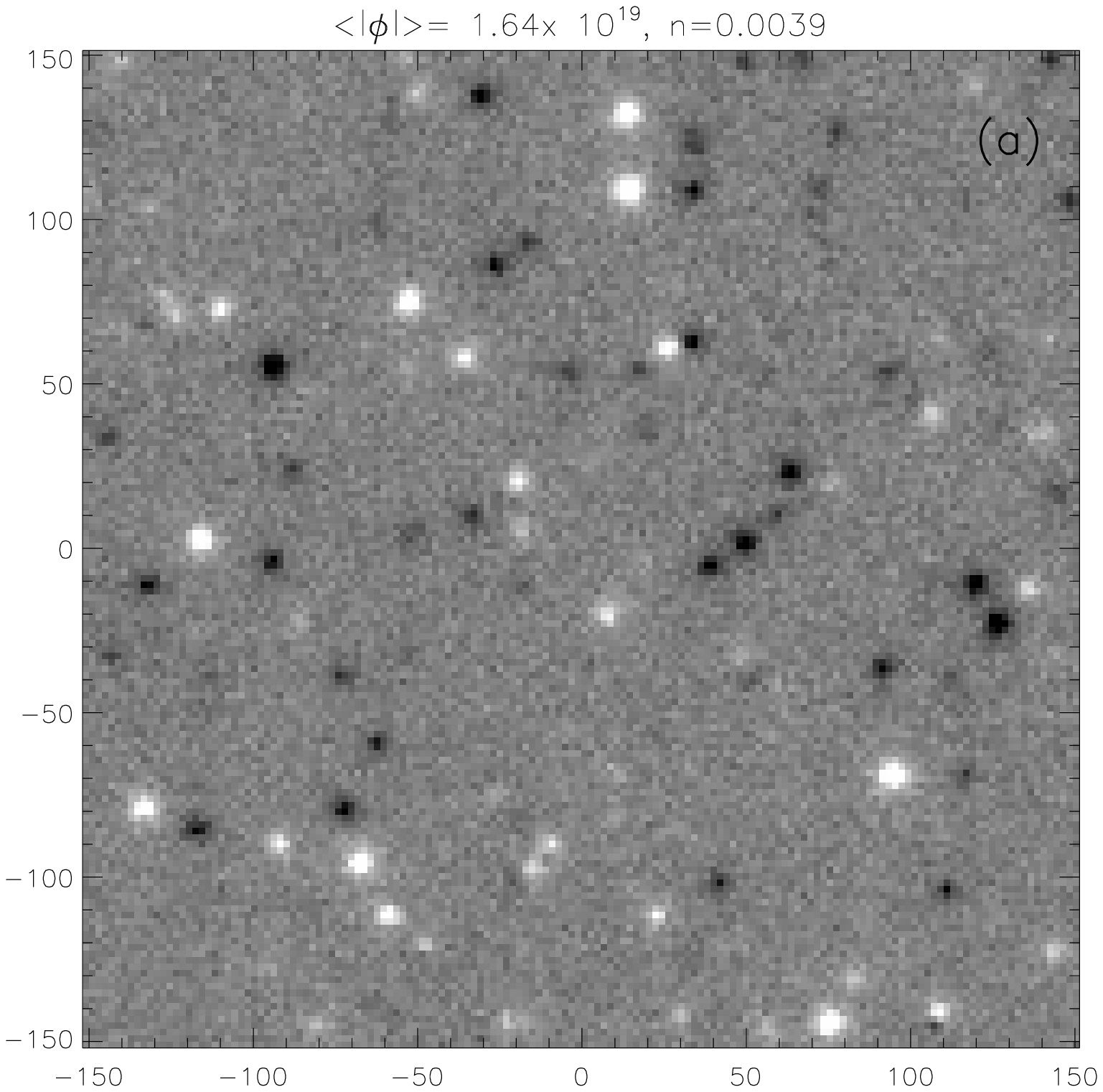,width=2.2in}%
\psfig{file=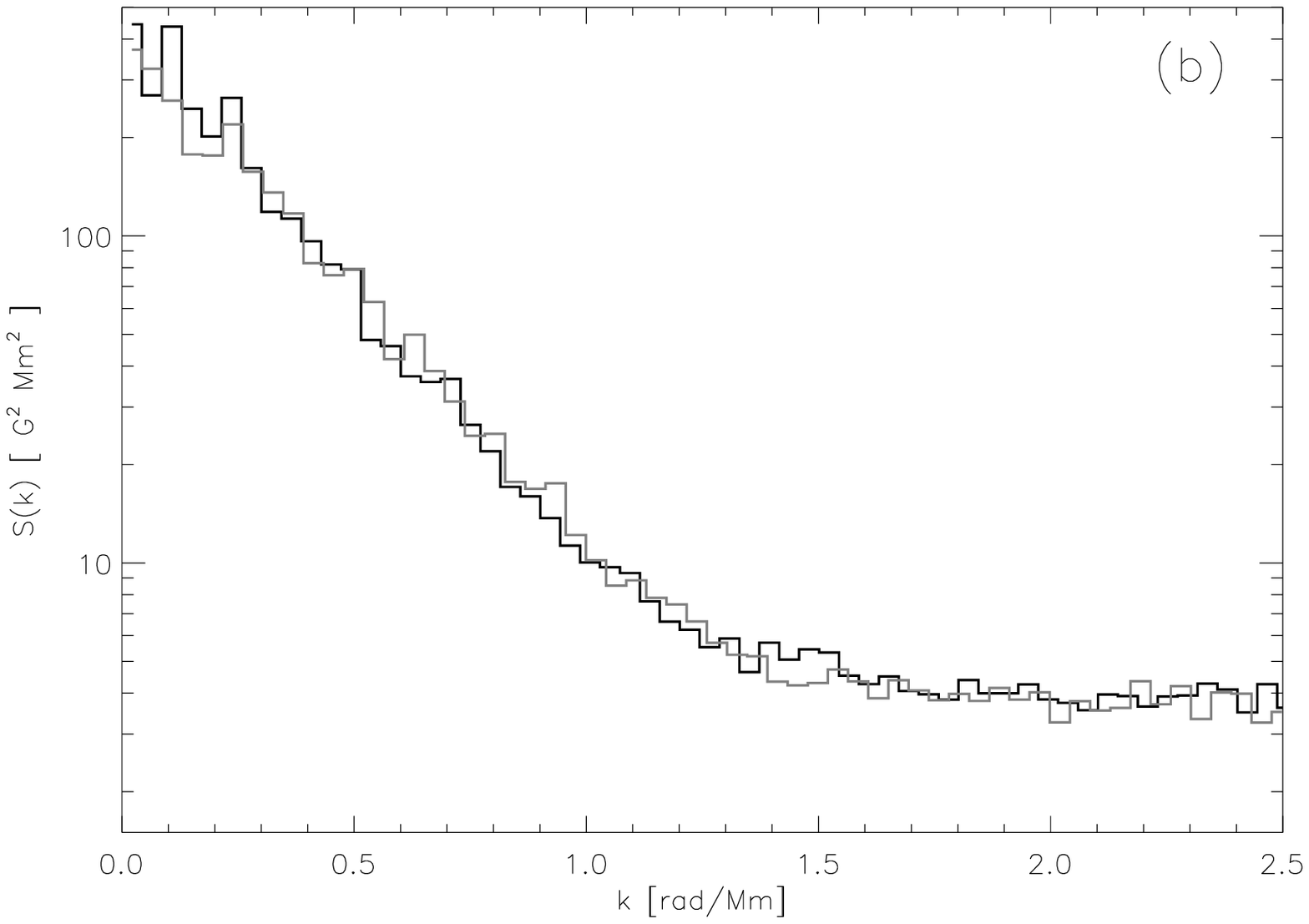,width=2.9in}}
\caption{A magnetogram synthesized from a distribution of point
sources at $d=2.2$ Mm below the photosphere.  (a) The magnetogram with
grey scale over $|B_z|<100$ G, exactly as for \ref{fig:mg}b.  
(b) The power spectra of this magnetogram (dark) and of \fig\
\ref{fig:mg}b (light).}
 	\label{fig:synth_cmp}
\end{figure}

Figure \ref{fig:synth_cmp}a shows a magnetogram synthesized from the
distribution of sources described above.  Its parameters were chosen
to resemble the five-minute average low resolution magnetogram of
\ref{fig:mg}b, and the two have indistinguishable power spectra
(\fig\ \ref{fig:synth_cmp}b).  The standard deviation and kurtosis of
the synthetic magnetogram, $\sigma_0=16.1$ G and $K_0=39$, are
well matched to the original.

Correcting the spectrum for an MTF with $\ell_0=1.6''$ leads to a
more narrowly peaked element, $f(r)$, whose radius is
$r_*=2.1$ Mm and which has $\chi=13.2$ 
(compared to $r_*=\sqrt{2}d=3.1$ Mm and
$\chi=6.4$ without correction; see \fig\ \ref{fig:shape}).
These more peaked elements must be distributed with
greater areal density and smaller mean flux 
in order to match the observed kurtosis and standard
deviation: $n=0.016\,{\rm Mm}^{-2}$ and 
$\avg{|\phi|}=5.6\times10^{18}$ Mx.
The instrument's point spread
function blurs each element reducing the total apparent variance.
The MTF correction thereafter amplifies the observed variance
leading us to the inference of a higher density of fundamental
elements.  In this case by a factor of four.

The same analysis can be performed on all of the spectra analyzed in
the previous section.  The MTF correction implies fundamental elements
with a median radius $r_*=1.1$ Mm, too small to observed directly in
low resolution.  As a consequence the median element density is quite
large, $n=0.046\,{\rm Mm}^{-2}$, and median flux quite small,
$\avg{|\phi|}=0.22\times10^{19}$ Mx.
These values are outside the range found by direct
identification of flux elements in magnetograms
\citep{Schrijver1997b,Parnell2002,DeForest2007} since the correction for
point spread function is not generally performed.  Indeed, our final
results depend quite sensitively on this correction, since
unresolved elements contribute only slightly to the statistics of
the observed magnetic field.

\begin{figure}[htp]
\psfig{file=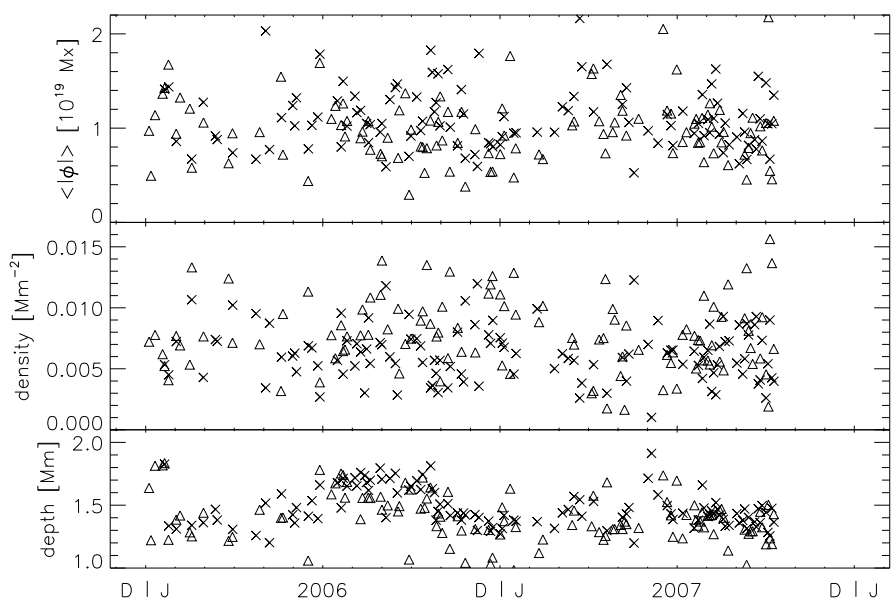,width=5.1in}
\caption{Parameters for the model magnetograms derived from spectra of
one-minute (triangles) and five-minute (crosses) full-disk magnetograms.
(top) The mean unsigned flux, $\avg{|\phi|}$, in units of $10^{19}$
Mx. (middle) The density of sources in units of Mm$^{-1}$.
(bottom) The depth $d$ at which the sources are placed.}
	\label{fig:phi_smry}
\end{figure}

We repeat our analysis without the MTF correction to obtain an
estimate which is at once more conservative and more consistent with
previous investigation. Figure \ref{fig:phi_smry} shows the 
depth (bottom panel), areal density, $n$ (middle panel) 
and flux, $\avg{|\phi|}$ (top panel) from low resolution
magnetograms over the 2006-2007 solar minimum.  
These exhibit the same slow variation as the
null density, presumably for the same reason: gradual variation of
focus.  The MTF-corrected version (not shown) has still more pronounced
variation due to its amplification of the high-wavenumber portion of
each spectrum.  The median depth is $d=1.43$ Mm, and
without the MTF correction the median values for areal
density and mean flux are $n=0.007\,{\rm Mm}^{-2}$ and
$\avg{|\phi|}=1.0\times10^{19}$ Mx respectively.  These are the value
for which the observations provide direct evidence and are
consistent with studies which directly identify flux  elements 
\citep{Schrijver1997b,Parnell2002,DeForest2007}.

\subsection{Using the model in a noise-free spectral estimate}

Submerged point sources alone, without the addition of white noise,
have the same purely exponential power spectrum as the
noise-corrected magnetograms.
It was found in \papi, however, that the
spectral estimate grossly overestimates the column of null points
above a distribution of submerged point sources.  This was 
attributed to a severe departure from gaussianity within a layer 
extending  $\Delta z\sim n^{-1/2}$ from the source plane 
($z\le15$ Mm in our case).  One
symptom of this departure is the divergence of the kurtosis given by
\eq\  (\ref{eq:kurt_mpole}).  The known overestimation for point 
sources raises concerns about the accuracy
of the spectral estimate from the noise-corrected magnetogram
spectrum, which has the same spectrum.

\begin{figure}[ht]
\centerline{\psfig{file=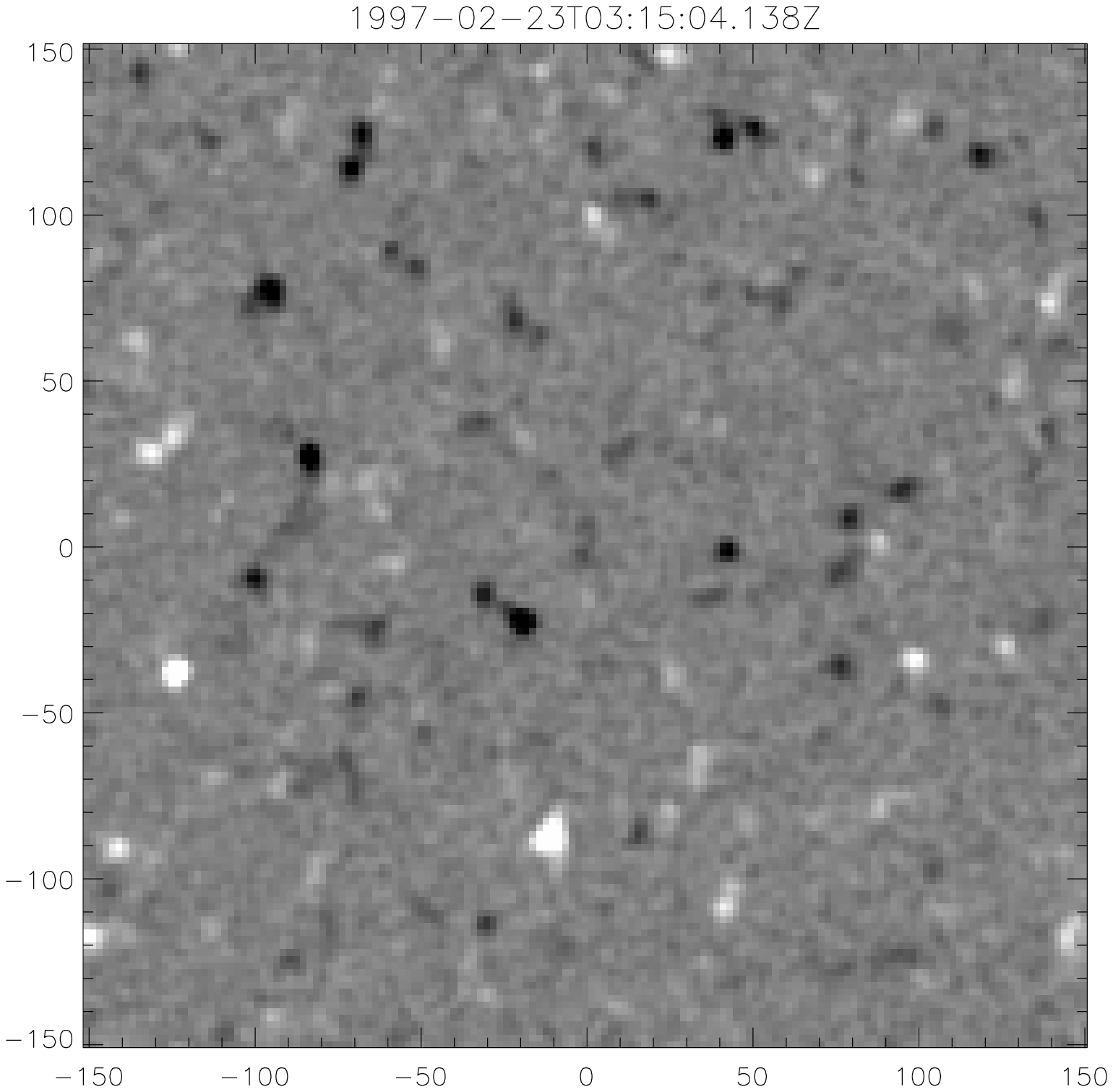,width=2.2in}%
\psfig{file=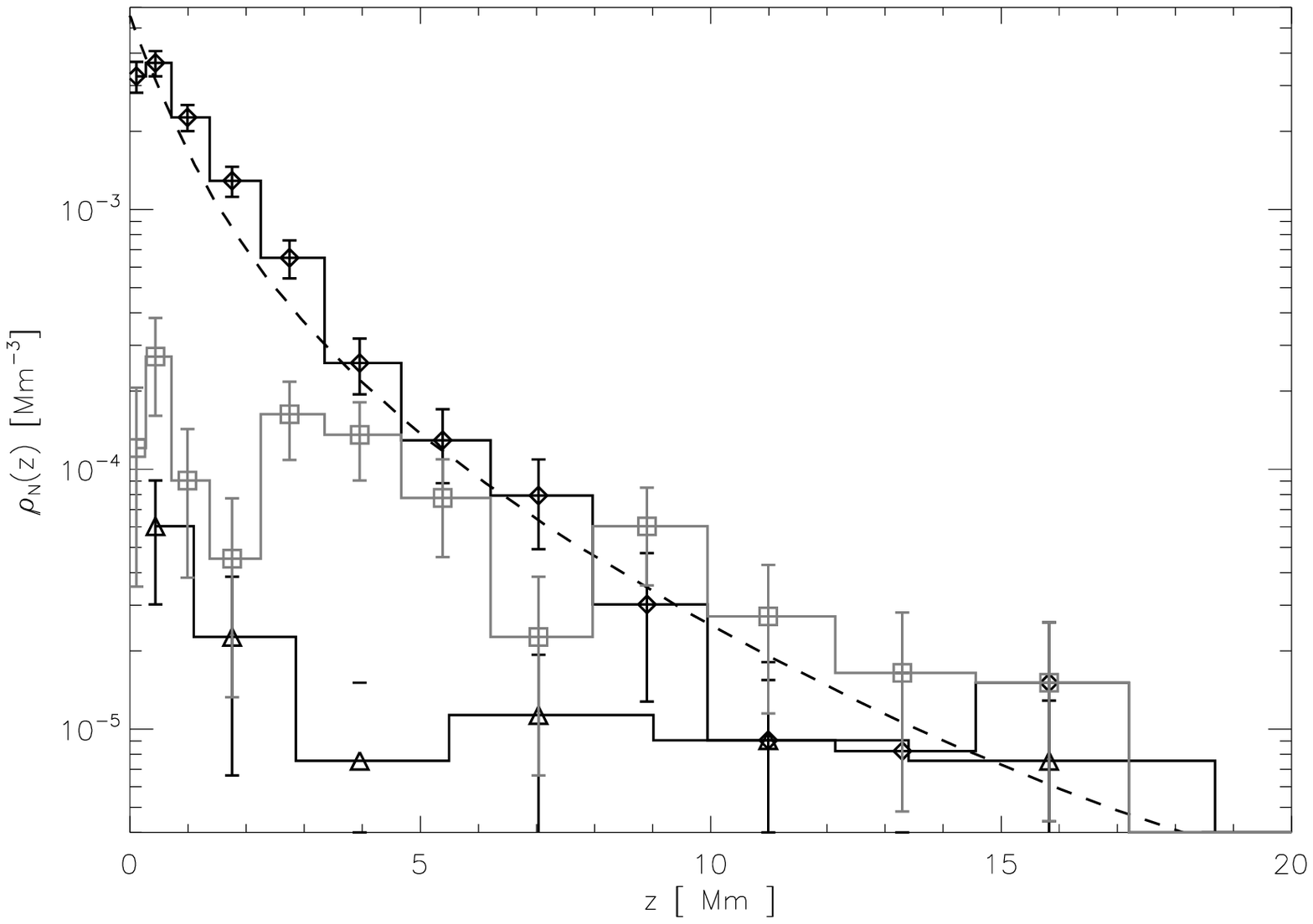,width=2.9in}}
\caption{The result of Fourier filtering the low-resolution 5-minute
average magnetogram \fig\ \ref{fig:mg}b.  (left) The magnetogram plotted
using the same grey scale as the original.  (right) The null densities,
$\rho_N(z)$, for the Fourier-filtered
magnetogram (diamonds) compared to the spectral
estimate (dashed).  The null densities from the synthetic 
magnetograms created with low and high densities 
of submerged point sources are plotted with triangles and squares
respectively.}
	\label{fig:clean}
\end{figure}

To further explore this comparison 
we reduced the noise in the low-resolution 5-minute
averaged magnetogram, \fig\ \ref{fig:mg}b, using a Fourier filter.   
Amplitudes of all high-spatial frequency Fourier modes
($|\kvec|>0.7\,{\rm rad/arcsec}$) were rescaled to fit
a pure exponential, $S(k)\propto e^{-2kd}$ with $d=2.2$, while
retaining their original phases. 
The result was a magnetogram (see \fig\ \ref{fig:clean}) which is
visibly cleaner, albeit blurrier than the original
(\fig\ \ref{fig:mg}b).  The central 
core of its distribution function fits a Gaussian of width
$\sigma^{(g)}_z=7.0$ G.  The original, 5-minute averaged magnetogram
had $\sigma^{(g)}_z=10.6$ G of which $k_c\sqrt{\pi S_n}=8.8$ G was
found to be white noise, leaving $\sigma_z=5.9$ G to be be genuine.
Thus it would seem our Fourier filtering removes a substantial amount
of the noise.

The Fourier-filtered magnetogram has, by construction, a psd
similar to a magnetogram synthesized from submerged
point sources (like \fig\ \ref{fig:synth_cmp}a without added
noise).  Such a magnetogram was synthesized, extrapolated
onto a grid and all its null points located
using the algorithm of Haynes and Parnell (2007).  The result was a
column density $N_n(1.5\,{\rm Mm})=0.3\times10^{-3}\,{\rm Mm}^{-2}$
of null points.
Repeating the procedure for the Fourier filtered magnetogram  yielded a
column, $N_n(1.5\,{\rm Mm})=3.6\times10^{-3}\,{\rm Mm}^{-2}$, more
than an order of magnitude greater.
The spectral estimates for both cases (which have similar
spectra) is $N_n(1.5\,{\rm Mm})=2.0\times10^{-3}\,{\rm Mm}^{-2}$; this
is smaller than the results from the previous section, which also
include MCT-correction.

The spectral estimate exceeds the column above point sources by
about a factor of seven, in agreement with the finding of \papi\
(they find an actual column $N_n(-d)=0.074n=
0.29\times10^{-3}\,{\rm Mm}^{-2}$).
On the other hand, it {\em underestimates} the column above the
filtered magnetogram by 44\%, in spite of its very large kurtosis:
$K_z=49.0$ (larger than the magnetogram before filtering).

To explore the significance of kurtosis, a 
second synthetic magnetogram was constructed with a greater areal 
density ($n=10^{-2}\,{\rm Mm}^{-2}$) of point 
sources at the same depth ($d=2.2$)
but with correspondingly smaller flux ($\avg{|\phi|}=0.69\times10^{19}$
Mx) to endow it with the same spectrum as the other two.  The kurtosis
of the resulting magnetogram ($K=29$) is lower than either of the
others.
A gridded extrapolation of this field was found to have a column 
$N_n(1.5\,{\rm Mm})=1.0\times10^{-3}\,{\rm Mm}^{-2}$.  Moreover, the
actual density, $\rho_N(z)$ falls well below the spectral estimate
below $z=n^{-1/2}-d=7.8$ Mm, as shown in the right panel of
\fig\ \ref{fig:clean}.  In
contrast, the null density of the Fourier-filtered magnetogram tracks
the spectral estimate, but is slightly above it, over the entire range
of heights.  The height at which the null density of the first
synthetic magnetogram becomes accurate, is $z=n^{-1/2}-d=13.8$ Mm,
consistent with \fig\ \ref{fig:clean}b.

These experiments suggest that the accuracy of the spectral estimate
in predicting the actual null point column, is unaffected by the
absence of white noise.  For both the filtered and unfiltered
5-minute magnetogram it underestimates the column by a similar factor.
This performance is far better than it would be if the 
photospheric field consisted of submerged point sources, as its
spectrum suggests.  The discrepancy cannot be explained by kurtosis of
the fields, since the spectral estimate is still too large for a
point-source-derived field with smaller kurtosis.  It seems that the
spectral estimate owes its success in the quiet Sun to a field
intrinsically more complex than one from point sources.

\section{Discussion}

The foregoing has shown that the Fourier spectra of quiet Sun
magnetograms have an invariant basic form, at least over moderately
high wave numbers ($0.2\,{\rm rad/arcsec}<k<1.5\,{\rm rad/arcsec}$).  
This spectral range plays the greatest role in
dictating the distribution of magnetic null points in the corona above.
As a consequence we find that the density of magnetic null
points in a potential field extrapolation is about 
$N_n(1.5\,{\rm Mm})=3.1\times 10^{-3}\,{\rm Mm}^{-2}$ 
over any patch of quiet Sun, varying by roughly $10\%$ from
day to day.  This value is just more than twice the
number of nulls found by R\'egnier, Parnell and Haynes (2008) 
at heights greater
than 1.5 Mm, using Hinode data of much higher resolution.  The discrepancy 
could possibly arise from the much greater
flux imbalance over that particular magnetogram.  The column we find
corresponds to 2 nulls above every 5 hexagonal, 14-Mm-diameter 
supergranules \citep{Hagenaar1997} .

The potential magnetic field extrapolated from a photospheric
magnetogram tends to become spatially smoother at increasing
heights.  This is a general property of harmonic functions and their
gradients.  One consequence is that the density of null points in such
a field decreases with height and most of the null points are found
close to the surface.  Of the null points above $z=1.5$ Mm
roughly $75\%$ are found within the next 3.0 Mm.
We find the spectral estimate of the null point
density in the generic quiet Sun to be
$\rho_n(z)\simeq0.04/(z+d)^{-3}$, where $d\simeq1.4\,{\rm Mm}$.  This
simple functional form follows from the approximately exponential
nature of the photospheric spectrum.

The density at very low heights, say $z<1\,{\rm Mm}$, is dictated by
very small horizontal structure in the photospheric field.  This
structure is the most challenging to observe, and its observation is
subject to the most instrumental distortion.  Moreover, this lowest
atmospheric layer, the chromosphere, has the greatest plasma pressure
and mass density and is least likely to satisfy the conditions which
would render a magnetic field potential, or even force free.  
For these reasons we choose to focus our investigation on null
points above $z=1.5\,{\rm Mm}$.  This nominal height has nothing to do
with the empirical depth, $d$, characterizing the exponential slope of
the photospheric spectra.  Rather, it arose from our experience with
different instrumentation (MDI low-resolution {\em versus} 
high-resolution) and different methods of accommodating instrumental 
effects
such as the MTF.  We found that the null point density, $\rho_N(z)$,
varied less above $z=1.5\,{\rm Mm}$ due to these instrumental
factors, so we deemed it to be more accurately measured there.

%  density of null points diminishes rapidly with height.
%    of nulls above 1.5 Mm, the vast majority are found in the 
%    next 3 Mm (75%).  the actual number depends critically on
%    m-gram spectra in range 0.3 rad/Mm - 1 rad/Mm.  scales
%    are adequately sampled by current instrumentation, MDI, but
%    requires careful attention.

The density and distribution
of null points in the corona is predicted somewhat reliably,
although not perfectly, by spectral estimate of Longcope,
Brown and Priest (2003, \papi).  Magnetic null points can be found
directly, and thus with more accuracy, in a potential field
extrapolated from a photospheric 
magnetogram.  We used this technique to test the spectral estimate,
and found it predicted the shape of the density,
$\rho_N(z)$, reasonably well.  The spectral estimate of the total column
was, however, too low by a factor of two to three above one Mm.  
If the spectral estimate for every magnetogram is low by the same 
factor, then the actual null 
density in the quiet Sun would be greater than the value we report.

This discrepancy is most 
likely due to a departure of the actual field from the statistical
properties assumed by the spectral estimate (homogeneity and
Gaussianity).  We did not find a satisfactory method of quantifying
these factors.  One characteristic in particular, the kurtosis of the
distribution, did not accurately predict when the assumptions were
adequately satisfied.  We did find, however, that the conditions
appear to be better satisfied, and the predictions better, for actual
magnetograms than for a random distribution of point magnetic sources.
In the latter case the spectral estimate {\em exceeds} 
the number of null points by at least an order of magnitude.

Although it is more accurate, direct enumeration of magnetic null
points suffers from several disadvantages relative to the spectral
estimate.  It is more computationally expensive, so performing the
search on over 500 different grids 
would be prohibitive.  More significantly, instrumental artifacts
such as noise and MTF are much more difficult to remove from the
extrapolation than to remove from the spectrum.  We found that white
noise present in MDI magnetograms can contribute artificial null points
outnumbering the real ones by as many as five-to-one.  The ease with
which such factors can be removed the Fourier spectra makes the
spectral estimate a superior one, even given its inaccuracy.

% Non-potential fields ?

We found that the spatial spectrum of the quiet Sun magnetic field is
well approximated as an exponential, at least for 
$k>0.2\,{\rm rad/Mm}$.  This spectrum is consistent with the potential
field from point magnetic sources submerged to a depth of $d\simeq1.4$
Mm.  These manifest themselves on the photosphere as circular
concentrations of flux, approximately $r_*=2.5$ Mm in radius.  We
are able to infer the size and areal density of these structures,
$\avg{|\phi|}=1.0\times10^{19}$ Mx and
$n=0.007\,{\rm Mm}^{-2}$, solely from the 
statistics of the photospheric field. The resolution of
such structures into finer elements, clearly seen in high-resolution 
observations, will affect the spectrum only at higher wavenumbers.  The
spectral form we observe suggests that such fine scales are
prone to cluster into 2.5 Mm concentrations.

Along similar lines, the imprint of structuring at still larger
scales, such as by the network, manifests itself at lower
wavenumbers, $k<0.2$ rad/Mm.  We find that actual quiet Sun spectra
differ the most over these low wavenumbers, suggesting more pronounced
day-to-day variation at these larger scales.  Furthermore, any particular
spectrum departs most from its exponential form over these low
wavenumbers.  This can be explained using fundamental elements 
distributed non-uniformly across the surface ---
concentrated, for example  along network boundaries.  This spectral
range dictates the distribution of null points at great heights, where
very few ever occur.  Thus the variation has virtually no effect on
the total column we report.

We performed our analysis on 562 different patches
($300''\times300''$) of quiet Sun
found at disk center over two different solar minima (1996-1998 and
2006-2007).  The magnetograms were chosen to be free of active
regions, plage or dominance of a single polarity.  No other data was
used to determine if the region was, for example, part of a coronal
hole.  Given their equatorial locations, however, it seems unlikely
that much coronal hole data was included in our survey.  We cannot
therefore say whether the null point density applies to coronal holes
as well as to equatorial quiet Sun.

These findings place in a clearer context observations and models with
putative relationships to coronal null points.  If the quiet Sun
corona is being heated primarily through dissipation at null
points, then the majority of this heat is supplied very low in the
corona.  The region between $z=1.5$ Mm and $z=4.5$ Mm contains three
times as many nulls as the whole space above it; it would therefore
probably receive a proportional share of the heating.  It remains to
be investigated whether the actual thermal structure of the quiet Sun
corona is consistent with this distribution of heating, and is
therefore consistent with heating by energy dissipation at null
points.

Along similar lines, if coronal jets were hypothesized to form
anywhere a coronal null point occurs, then our results can be used in
an {\em ab initio} 
prediction of their frequency.  Moreover, the distribution of their cusp
heights should be somehow influenced by the distribution of null
heights: short jets should out-number tall ones.  Actual jets
undoubtedly involve many factors, in addition to magnetic
geometry, which influence the distribution of their properties.  Still,
it would be difficult to invoke null points as a 
requirement for jets if jets of a given height outnumbered null points
at that level.  A survey of coronal jets comprehensive enough to
determine their frequency would thereby cast some light on the
importance of null points in their construction.

\appendix

\section{Estimating the modulation transfer function (MTF)}

To estimate the MTF, consider an ideal imaging system 
characterized by the same point spread function at every point on the
image plane (i.e.\ an isoplanatic system).
In this case the power spectral density (psd), $S(k)$, of the image 
is the product of the actual psd, $\tilde{S}(k)$, and the square of
the modulation transfer function (MTF), $M(k)$ \citep{Born1980}.  
The MTF is the magnitude of the Fourier transform of the normalized point
spread function.  Generally, $M(k)$, is a decreasing function of
wave-number, causing the observed psd, $S(k)$, 
to be steeper than the actual one, $\tilde{S}(k)$.

If the same field is observed by MDI in both its
high and low resolution modes the
underlying psd, $\tilde{S}(k)$, will be the same for both images.  In
this case
\be
  {S_{\rm lo}(k)\over S_{\rm hi}(k)} ~=~
  {M^2_{\rm lo}(k)\over M^2_{\rm hi}(k)} ~~.
  \label{eq:mtf_rat}
\ee
The MTF of the low resolution images will be dominated by the
defocussing intended to reduce aliasing from
spatial frequencies above the Nyquist limit
$k_{n}=1.6$ rad/arcsec \citep{Scherrer1995}.
It is this contribution we can ascertain by cross comparison.

One-minute high resolution magnetograms were obtained for 11 different
intervals on five different dates throughout 2007 while MDI was in
focus position 5.  
A single psd, $S_{\rm hi}(k)$, was constructed for each time
interval.  In cases where several (typically 4) consecutive one-minute
magnetograms were available from a given interval, the individual psd's
were averaged together to produce $S_{\rm hi}(k)$ with lower noise.  The
nearest 96-minute full disk low-resolution magnetogram was then used
to compute $S_{\rm lo}(k)$.  The high and low resolution magnetograms
were separated by six to 30 minutes (typically 10 minutes)
strengthening the assumption that $\tilde{S}(k)$ is the same for both
observations.

All psd's were computed as described in section \ref{sec:psd} using
spectral annuli of width $\Delta k=0.05$ rad/arcsec, 
and the white noise floor was
subtracted from each psd.  The ratio of \eq\ (\ref{eq:mtf_rat}) was
computed over the range $k<1.5$ rad/arcsec where both spectra 
were available and fairly free of noise.  Full 
disk magnetograms have been re-calibrated by the
MDI instrument team while high-resolution magnetograms 
have not.  To compensate for the different calibration we divided the
ratio by the mean of its lowest 8 spectral bins ($k<0.4$ rad/arcsec).

\begin{figure}[htp]
\centerline{\psfig{file=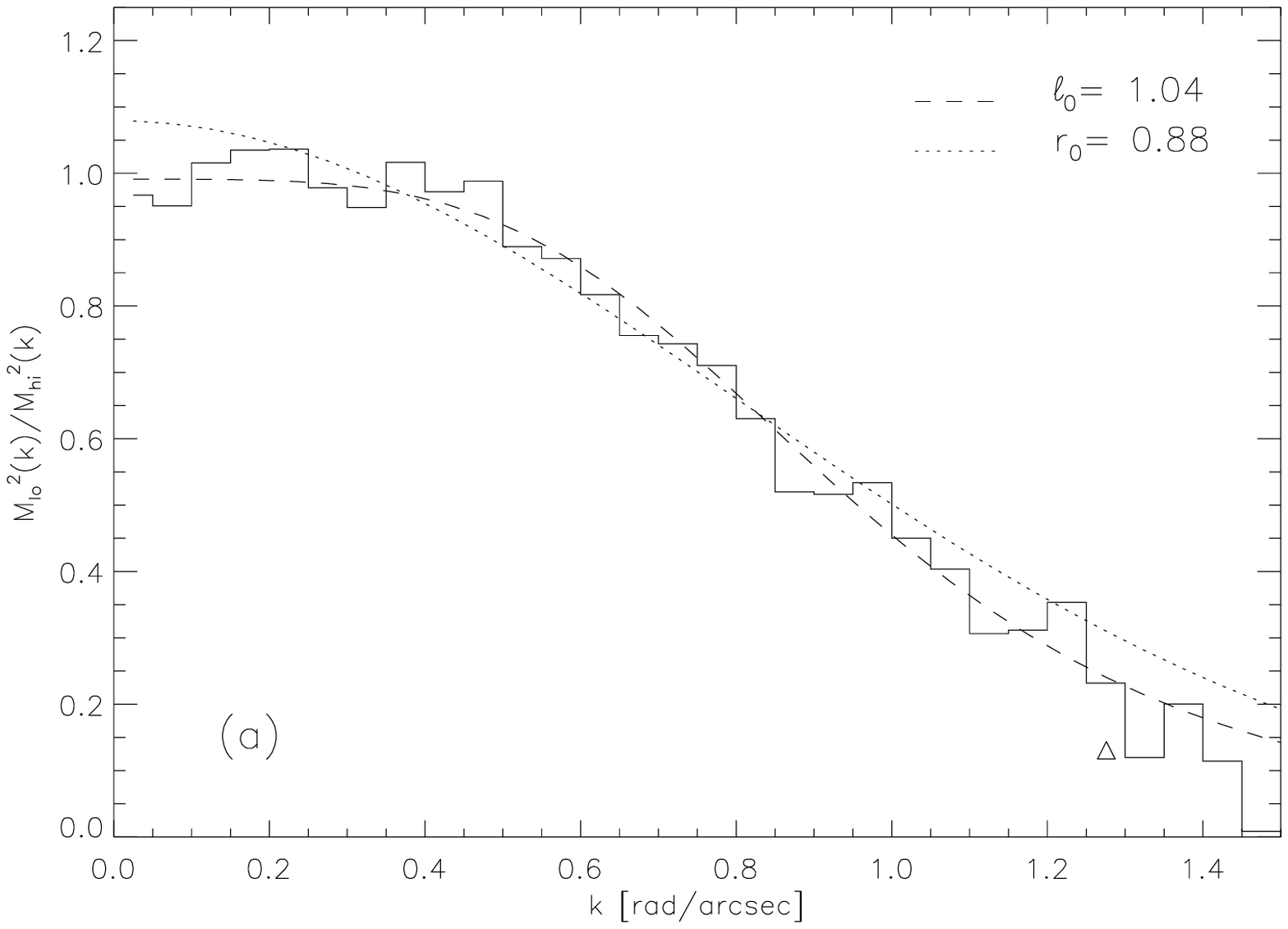,width=2.7in}%
\psfig{file=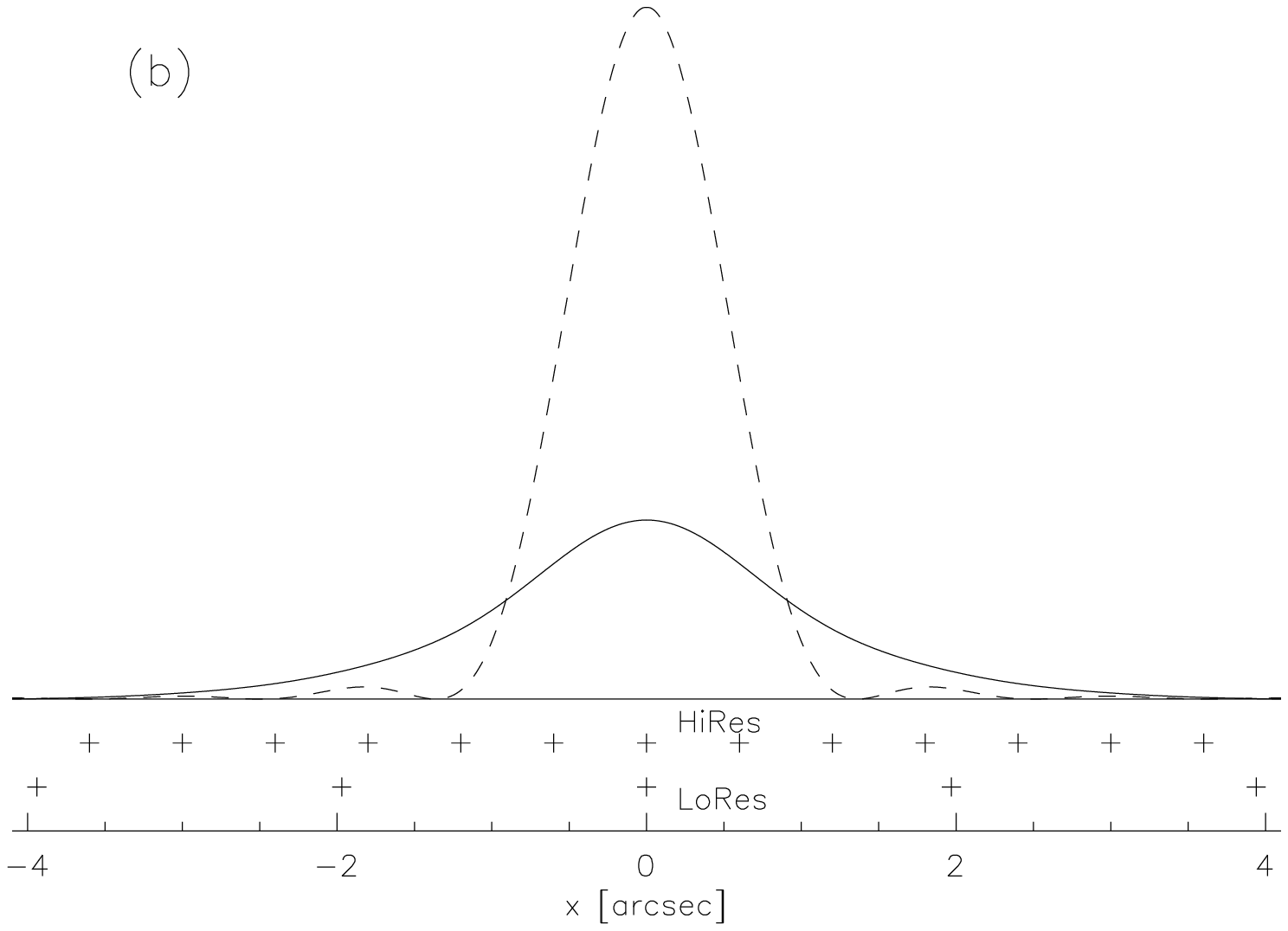,width=2.4in}}
\caption{The MTF and point spread function used for MDI.  (a)
The median of MTF estimates from 11 quiet Sun observations.
The dark solid curve is the estimate from the ratio of \eq\
(\ref{eq:mtf_rat}).  Dashed and dotted curves are empirical fits.
The triangle is the pre-launch value of the ratio quoted by
Scherrer \etal\ (1995).
(b) The point spread functions for the high
resolution (dashed) and low resolution (solid) based on the MTF
calculation.  Sample pixel locations for each resolution are shown
below the curves with crosses.}
                     \label{fig:mtf_est}
\end{figure}

A median MTF ratio is constructed by taking the median value of 
$S_{\rm lo}(k)/S_{\rm hi}(k)$ from 
all 11 time intervals for each spectral bin (see \fig\
\ref{fig:mtf_est}a).  It is fairly flat over the long-wavelength range
and decays to low values by $k>1.2$ rad/arcsec.  Although the low resolution
magnetograms extend out to $k_c = 1.8$ rad/arcsec, it appears there is little
sensitivity over that last third of that range.

\nocite{Scherrer1995}

The estimated MTF ratio 
from \fig\ \ref{fig:mtf_est}a was fitted, over the range
$0.1\,{\rm rad/arcsec}<k<1\,{\rm rad/arcsec}$, 
by a variety of empirical functions.  Dashed and dotted
curves show two of the more successful functions
\begin{eqnarray}
  M^2_{\rm lo}/M^2_{\rm hi} &=& {A\over 1 + \ell_0^4 k^4} ~~,
  \label{eq:MTF_empirical} \\
  M^2_{\rm lo}/M^2_{\rm hi}  &=& A' e^{-r_0^2k^2} ~~,
  \label{eq:MTF_gauss}
\end{eqnarray}
respectively.  The best fits were for parameter values 
$\ell_0=1.04''$ and
$r_0=0.88''$.   The former curve better fits the  flat inner
region and the steeper fall-off beyond $k\simeq0.6$ rad/arcsec.
There is evidence that the actual MTF falls even
faster, perhaps even reaching zero, for $k>1.4$ rad/arcsec.  Dividing the
observed spectrum by expression (\ref{eq:MTF_empirical}) is therefore
a conservative correction at the very highest wave numbers.

While this cross comparison provides the ratio of the two MTFs, it
cannot give any information about either one alone.  The most conservative
correction to either spectrum is to assume the high-resolution
mode is diffraction limited and in perfect focus.  In this case its
MTF would be \citep{Ghatak1978}
\be
  M_{\rm hi}(k) ~=~ M_d(k) ~=~ {2\over\pi}\left[\, 
  \cos^{-1}\left({k\over k_0}\right) ~-~
  {k\over k_0}\sqrt{1 - {k^2\over k_0^2}} \, \right] ~~,
	\label{eq:MTF_d}
\ee
where $k_0=2\pi d/\lambda_{\rm Ni} = 
5.63\,{\rm rad/arcsec}$ for a primary mirror of diameter, $d=12.5$ cm 
and optical wavelength $\lambda_{\rm Ni}=6768\AA$ \citep{Scherrer1995}.

\nocite{Scherrer1995} 
Scherrer \etal\ (1995) reported that the intentional 
defocussing reduced the low resolution MTF at $k=0.8k_n=1.3$ rad/arcsec 
to $M=0.17$ from its diffraction-limited value of $M_d=0.47$.  That is
to say $M_{\rm lo}/M_{\rm hi}=0.36$, under our assumption that 
$M_{\rm hi}=M_d$.  The square of this value is
plotted as a triangle in \fig\ \ref{fig:mtf_est}.  Our data point is
consistent with this value, but the empirical fits are greater at this
end of the spectrum.  Using the empirical fits will then under-correct
the highest wave-numbers.

\acknowledgements

We wish to thank Phil Scherrer, Todd Hoeksema, Jesper Schou 
and Craig DeForest for helpful discussions concerning MDI.  This 
work was supported in part by NASA's TR\&T program.

%\bibliography{/home/dana/tex/inputs/short_abbrevs,/home/dana/tex/full_lib}
%\bibliography{/Users/danalongcope/stuff//long_abbrevs.bib,/Users/danalongcope/stuff/full_lib.bib}

\begin{thebibliography}{36}
\providecommand{\natexlab}[1]{#1}

\bibitem[\protect\citeauthoryear{Abramenko \emph{et~al.}}{2001}]{Abramenko2001}
Abramenko, V., Yurchyshyn, V., Wang, H., Goode, P.~R.: 2001,
  \emph{Solar~Phys.} \textbf{201}, 225.

\bibitem[\protect\citeauthoryear{Antiochos}{1998}]{Antiochos1998}
Antiochos, S.~K: 1998, \emph{Astrophys. J} \textbf{502}, L181.

\bibitem[\protect\citeauthoryear{{Beveridge}, Priest and 
Brown}{2004}]{Beveridge2004}
{Beveridge}, C., {Priest}, E.~R., {Brown}, D.~S.: 2004, \emph{Geophysical and
  Astrophysical Fluid Dynamics} \textbf{98}, 429.

\bibitem[\protect\citeauthoryear{Born and Wolf}{1980}]{Born1980}
Born, M., Wolf, E.: 1980, \emph{Principles of Optics: Electromagnetic theory
  of propagation and diffraction of light}, Pergamon Press, New York, NY, sixth
  edition.

\bibitem[\protect\citeauthoryear{{Cirtain} \emph{et~al.}}{2007}]{Cirtain2007}
{Cirtain}, J.~W., {Golub}, L., {Lundquist}, L., {van Ballegooijen}, A.,
  {Savcheva}, A., {Shimojo}, M., {DeLuca}, E., {Tsuneta}, S., {Sakao}, T.,
  {Reeves}, K., {Weber}, M., {Kano}, R., {Narukage}, N., {Shibasaki}, K.:
  2007, \emph{Science} \textbf{318}, 1580.

\bibitem[\protect\citeauthoryear{Close, Parnell and Priest}{2004}]{Close2004b}
Close, R.~M., Parnell, C.~E., Priest, E.~R.: 2004, \emph{Solar~Phys.}
  \textbf{225}, 21.

\bibitem[\protect\citeauthoryear{{Craig}, Fabling and Henton}{1995}]{Craig1995}
{Craig}, I.~J.~D., {Fabling}, R.~B., {Henton}, S.~M., {Rickard}, G.~J.: 1995,
  \emph{Astrophys. J} \textbf{455}, L197.

\bibitem[\protect\citeauthoryear{Craig  and Mc{C}lymont}{1993}]{Craig1993}
Craig, I. J.~D., Mc{C}lymont, A.~N.: 1993, \emph{Astrophys. J} \textbf{405},
  207.

\bibitem[\protect\citeauthoryear{{Culhane} \emph{et~al.}}{2007}]{Culhane2007}
{Culhane}, L., {Harra}, L.~K., {Baker}, D., {van Driel-Gesztelyi}, L., {Sun},
  J., {Doschek}, G.~A., {Brooks}, D.~H., {Lundquist}, L.~L., {Kamio}, S.,
  {Young}, P.~R., {Hansteen}, V.~H.: 2007, \emph{Pub. Astron. Soc. Japan} \textbf{59}, 751.

\bibitem[\protect\citeauthoryear{DeForest \emph{et~al.}}{2007}]{DeForest2007}
DeForest, C.~E., Hagenaar, H.~J., Lamb, D.~A., Parnell, C.~E., Welsch, B.~T.:
  2007, \emph{Astrophys. J} \textbf{666}, 576.

\bibitem[\protect\citeauthoryear{Dungey}{1958}]{Dungey1958}
Dungey, J.~W.: 1958, \emph{Cosmic Electrodynamics}, Cambridge University Press,
  Cambridge, U.K.

\bibitem[\protect\citeauthoryear{Galsgaard and Nordlund}{1997}]{Galsgaard1997}
Galsgaard, K., Nordlund, A.: 1997, \emph{J Geophys. Res.} \textbf{102}, 231.

\bibitem[\protect\citeauthoryear{Ghatak and Thyagarajan}{1978}]{Ghatak1978}
Ghatak, A.~J., Thyagarajan, K.: 1978, \emph{Contemporary Optics}, Plenum
  Press, New York.

\bibitem[\protect\citeauthoryear{Greene}{1988}]{Greene1988}
Greene, J.~M.: 1988, \emph{J Geophys. Res.} \textbf{93}, 8583.

\bibitem[\protect\citeauthoryear{Hagenaar, Schrijver and Title}{1997}]{Hagenaar1997}
Hagenaar, H.~J., Schrijver, C.~J., Title, A.~M.: 1997, \emph{Astrophys. J}
  \textbf{481}, 988.

\bibitem[\protect\citeauthoryear{Hassam}{1992}]{Hassam1992}
Hassam, A.~B.: 1992, \emph{Astrophys. J} \textbf{399}, 159.

\bibitem[\protect\citeauthoryear{Haynes and  Parnell}{2007}]{Haynes2007}
Haynes, A.~L.,  Parnell, C.~E.: 2007, \emph{Phys.~Plasmas} \textbf{14}, 2107.

\bibitem[\protect\citeauthoryear{Hesse and Schindler}{1988}]{Hesse1988}
Hesse, M., Schindler, K.: 1988, \emph{J Geophys. Res.} \textbf{93}, 5559.

\bibitem[\protect\citeauthoryear{Hufnagel and Stanley}{1964}]{Hufnagel1964}
Hufnagel, R.~E., Stanley, N.~R.: 1964, \emph{J. Optical Soc. America} 52--61.

\bibitem[\protect\citeauthoryear{Liu and Norton}{2001}]{Liu2001}
Liu, Y., Norton, A.~A.: 2001, Mdi measurement errors: The magnetic
  perspective, Technical Report SOI Technical Note 01-144, Standford SOI.

\bibitem[\protect\citeauthoryear{Longcope, Brown and Priest}{2003}]{Longcope2003b}
Longcope, D.~W., Brown, D.~S., Priest, E.~R.: 2003, \emph{Phys.~Plasmas}
  \textbf{10}, 3321.

\bibitem[\protect\citeauthoryear{{McLaughlin} and {Hood}}{2004}]{McLaughlin2004}
{McLaughlin}, J.~A., {Hood}, A.~W.: 2004, \emph{Astron. Astrophy.}
  \textbf{420}, 1129.

\bibitem[\protect\citeauthoryear{{Moreno-Insertis}, Galsgaard and Ugarte-Urra}{2008}]{MorenoInsertis2007}
{Moreno-Insertis}, F., {Galsgaard}, K., {Ugarte-Urra}, I.: 2008,
  \emph{Astrophys. J} \textbf{673}, L211.

\bibitem[\protect\citeauthoryear{Parnell}{2002}]{Parnell2002}
Parnell, C.: 2002, \emph{Mon. Not. Roy. Astron. Soc.} \textbf{335}, 389.

\bibitem[\protect\citeauthoryear{{Pontin} and {Galsgaard}}{2007}]{Pontin2007}
{Pontin}, D.~I., {Galsgaard}, K.: 2007, \emph{J Geophys. Res.} \textbf{112},
  3103.

\bibitem[\protect\citeauthoryear{Press \emph{et~al.}}{1986}]{Press1986}
Press, W.~H., Flannery, B.~P., Teukolsky, S.~A., Vetterling, W.~T.: 1986,
  \emph{Numerical Recipes: The art of scientific computing}, Cambridge
  University Press, Cambridge.

\bibitem[\protect\citeauthoryear{R{\'e}gnier, Parnell and Haynes}{2008}]{Regnier2008}
R{\'e}gnier, S., Parnell, C.~E., Haynes, A.~L.: 2008, \emph{Astron. Astrophy.}
  \textbf{484}, L47.

\bibitem[\protect\citeauthoryear{{Rickard} and {Titov}}{1996}]{Rickard1996}
{Rickard}, G.~J., {Titov}, V.~S.: 1996, \emph{Astrophys. J} \textbf{472}, 840.

\bibitem[\protect\citeauthoryear{Scherrer \emph{et~al.}}{1995}]{Scherrer1995}
Scherrer, P.~H., Bogart, R.~S., Bush, R.~I., Hoeksema, J.~T., Kosovichev,
  A.~G., Schou, J., Rosenberg, W., Springer, L., Tarbell, T.~D., Title, A.,
  Wolfson, C.~J., Zayer, I.,   {MDI Engineering Team}: 1995, \emph{Solar~Phys.}
  \textbf{162}, 129.

\bibitem[\protect\citeauthoryear{Schrijver and Title}{2002}]{Schrijver2002}
Schrijver, C.~J.,  Title, A.~M.: 2002, \emph{Solar~Phys.} \textbf{207}, 223.

\bibitem[\protect\citeauthoryear{Schrijver \emph{et~al.}}{1997}]{Schrijver1997b}
Schrijver, C.~J., Title, A.~M., Hagenaar, H.~J., Shine, R.~A.: 1997,
  \emph{Solar~Phys.} \textbf{175}, 329.

\bibitem[\protect\citeauthoryear{Seehafer}{1986}]{Seehafer1986}
Seehafer, N.: 1986, \emph{Solar~Phys.} \textbf{105}, 223.

\bibitem[\protect\citeauthoryear{{Shibata} \emph{et~al.}}{1992}]{Shibata1992}
{Shibata}, K., {Ishido}, Y., {Acton}, L.~W., {Strong}, K.~T., {Hirayama}, T.,
  {Uchida}, Y., {McAllister}, A.~H., {Matsumoto}, R., {Tsuneta}, S., {Shimizu},
  T., {Hara}, H., {Sakurai}, T., {Ichimoto}, K., {Nishino}, Y., {Ogawara},
  Y.: 1992, \emph{Pub. Astron. Soc. Japan} \textbf{44}, L173.

\bibitem[\protect\citeauthoryear{{Shimojo} \emph{et~al.}}{1996}]{Shimojo1996}
{Shimojo}, M., {Hashimoto}, S., {Shibata}, K., {Hirayama}, T., {Hudson}, H.~S.,
 {Acton}, L.~W.: 1996, \emph{Pub. Astron. Soc. Japan} \textbf{48}, 123.

\bibitem[\protect\citeauthoryear{Sweet}{1958}]{Sweet1958}
Sweet, P.~A.: 1958, in B.~Lehnert (ed.), \emph{Electromagnetic Phenomena in
  Cosmical Physics}, 123--134, Cambridge University Press, Cambridge, U.K.

\bibitem[\protect\citeauthoryear{Yokoyama and Shibata}{1996}]{Yokoyama1996}
Yokoyama, T., Shibata, K.: 1996, \emph{PASJ} \textbf{48}, 353.

\end{thebibliography}

%\end{article}
\end{document}